\documentclass[prd,preprint,showpacs,preprintnumbers,nofootinbib,eqsecnum,superscriptaddress,longbibliography]{revtex4-1}

 \usepackage[dvips,final]{graphicx}
  \usepackage{amssymb}
   \usepackage{amsmath}
    \usepackage{amsfonts}
     \usepackage{epsfig}
      \usepackage{bm}

\usepackage{mathpazo}

\usepackage[section]{placeins}

\usepackage{multirow}
\usepackage{booktabs}
\usepackage{array}
\usepackage{tabularx}
\usepackage{xcolor}
\usepackage{pstricks}

\def\lsim{\mathrel{\rlap{\lower4pt\hbox{\hskip1pt$\sim$}}
    \raise1pt\hbox{$<$}}}         
\def\gsim{\mathrel{\rlap{\lower4pt\hbox{\hskip1pt$\sim$}}
    \raise1pt\hbox{$>$}}}         

\newcommand{\br}{\mbox{\boldmath $r$}}

\newcommand{\bk}{\mbox{\boldmath $k$}}
\newcommand{\bDelta}{\mbox{\boldmath $\Delta$}}
\newcommand{\bPhi}{\mbox{\boldmath $\Phi$}}
\newcommand{\bkappa}{\mbox{\boldmath $\kappa$}}

\newcommand{\be}{\mbox{\boldmath $e$}}

\newcommand{\half}{{1\over 2}}

\begin{document}

\title{
Exclusive production of $\rho$ meson in gamma-proton collisions: $d \sigma/dt$ and the role of helicity flip processes
}

\author{Anna Cisek}
\email{acisek@ur.edu.pl}
\affiliation{College of Natural Sciences, Institute of Physics, University of Rzesz\'ow,
ul. Pigonia 1, PL-35-310 Rzesz\'ow, Poland}

\author{Wolfgang Sch\"afer}
\email{Wolfgang.Schafer@ifj.edu.pl}
\affiliation{Institute of Nuclear Physics, Polish Academy of Sciences,
ul. Radzikowskiego 152, PL-31-342 Krak{\'o}w, Poland}

\author{Antoni Szczurek}
\email{Antoni.Szczurek@ifj.edu.pl}
\affiliation{Institute of Nuclear Physics, Polish Academy of Sciences,
ul. Radzikowskiego 152, PL-31-342 Krak{\'o}w, Poland}
\affiliation{College of Natural Sciences, Institute of Physics, University of Rzesz\'ow,
ul. Pigonia 1, PL-35-310 Rzesz\'ow, Poland}

\begin{abstract}
We calculate the differential cross section $d\sigma/dt$ for the diffractive photoproduction process $\gamma p \to \rho p$ and compare to recent data extracted by the CMS collaboration from ultraperipheral proton-lead collisions. Our model is based on two-gluon exchange in the nonperturbative domain.
We take into account both helicity conserving and often neglected helicity-flip
amplitudes in the $\gamma \to V$ transition.
The letter can contribute at finite $t$.
The shape of the differential cross section as well as the role of
helicity flip processes is strongly related to the dependence of 
the unintegrated gluon distribution on transverse momenta in 
the nonperturbative region.
\end{abstract}
\maketitle

\section{Introduction}

The exclusive photoproduction of vector mesons is one of the
intensively studied processes at high energies. 
It can be seen as a diffractive excitation of a photon into a vector meson state. Being a diffractive process it shows a prominent forward peak, and within the forward cone $s$-channel helicity is expected to be conserved to very good accuracy. For the light vector mesons, like the $\rho$ which we discuss in this work, the energy dependence displays a ``soft pomeron'' behaviour and follows the one of the total $\gamma p$ photoabsorption cross section. 
At high $\gamma^* p$ cm-energies and large virtualities $Q^2$ of the photon, 
a QCD factorization approach based on a twist expansion of meson distribution amplitudes and unintegrated gluon distributions (UGDs) ~\cite{Anikin:2009bf} can be used to constrain the gluon UGD ~\cite{Bolognino:2018rhb,Bolognino:2021niq}.
On the other hand in the photoproduction limit, $Q^2 = 0$ many theoretical approaches to diffractive $\rho$-meson production are based on the vector meson dominance ansatz and essentially describes the process as an elastic scattering of a $\rho$-meson on the proton.
Besides being problematic in the continuation to large $Q^2 \gg m_\rho^2$, in this approach also the helicity and transverse momentum dependence of the amplitude have to be inserted ``by hand''. Indeed, within the diffraction cone, models of soft diffractive processes often impose the $s$-channel helicity conservation, following \cite{Gilman:1970vi}.

Our work is motivated by a  recent measurement of the differential cross section $d \sigma/dt$ \cite{CMS:2018ixo}. This observable has been advocated as a probe of gluon saturation effects \cite{Armesto:2014sma, Goncalves:2018blz}. These calculations, which are formulated in the color dipole approach also restrict themselves to the helicity conserving part of the amplitude.
In this work we wish to investigate the shape of the differential cross section,
some questions that come to mind are: do we expect dips in the differential cross section, or what is the role of spin-flip transitions at intermediate and large $t$? We will see, that the answer to these questions is indeed strongly dependent
on the modelling of the unintegrated glue.
We compare our calculations for a variety of unintegrated gluon distributions available in the literature, and compare to the recent (still preliminary) data of the CMS collaboration \cite{CMS:2018ixo} which were obtained from the ultraperipheral $p Pb$-collisions. 

 \section{Sketch of the formalism}
 
\begin{figure}
  \centering
  \includegraphics[width=.8\textwidth]{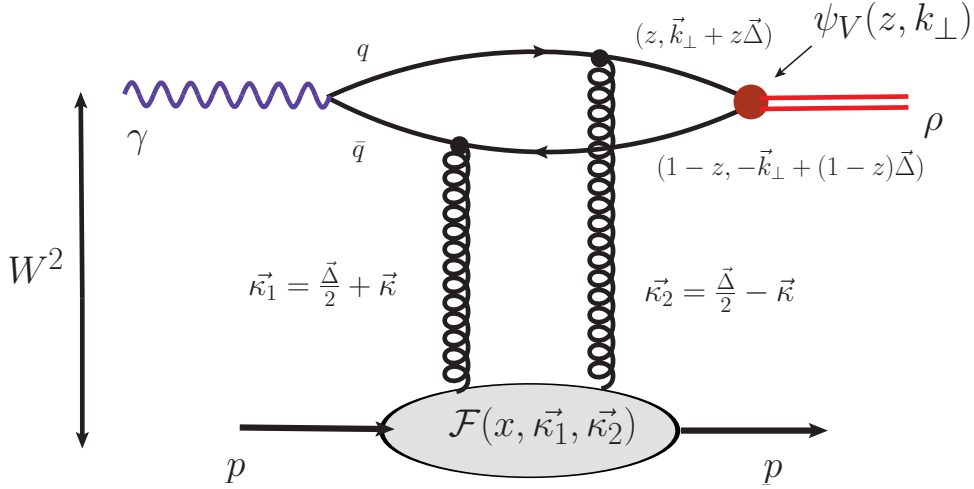}
  \caption{Sample Feynman diagram for the $\gamma p \to \rho p$ diffractive amplitude.}
\label{fig:diagram}
\end{figure}
The imaginary part of the amplitude for the reaction 
$\gamma(\lambda_\gamma) p \to V(\lambda_V ) p$ depicted in Fig.~\ref{fig:diagram} can be written 
as \cite{Ivanov:2004ax}:
\begin{eqnarray}
\Im m \, {\cal M}_{\lambda_V,\lambda_\gamma}(W,\bDelta) &=&
W^2 \frac{c_{V} \sqrt{4 \pi \alpha_{em}}}{4 \pi^2} \, 
\int 
{d\kappa^2 \over \kappa^4} \alpha_{S}(q^2)  {\cal{F}}(x,{\bDelta \over 2} +\bkappa,{\bDelta \over 2} - \bkappa) \nonumber \\
&\times&\int \frac{dzd^{2}\bk}{z(1-z)} I(\lambda_{V},\lambda_{\gamma};z, \bkappa,\bk,\bDelta) \psi_V(z,k) \; .
\end{eqnarray}
Here $W$ is the $\gamma p$-cm energy, $\bDelta$ is the transverse momentum transfer, and for the $\rho$-meson $c_V=1/\sqrt{2}$. 
Furthermore, $\psi_V(z,k)$ is a (radial part of the) light cone wave funtion of the vector meson, and ${\cal F}(x,\bkappa_1,\bkappa_2)$ is a generalization of the unintegrated gluon distribution valid off the forward direction.
Its first argument has to be understood as $x= M_V^2/W^2$, where $M_V$ is the mass of the vector meson. 

The impact factors $I(\lambda_V,\lambda_\gamma)$  beyond $s$-channel helicity conservation have been obtained in \cite{Ivanov:1998gk,Kuraev:1998ht} and are conveniently summarized in \cite{Ivanov:2004ax}.

We are interested in the case of real, transversely polarized photons, so that the following amplitudes depending on the helicities $\lambda_\gamma$ and $\lambda_V$ of photon and meson, contribute:
\begin{enumerate}
    \item The $s$-channel helicity conserving $T \to T$ transition, where $\lambda_\gamma = \lambda_V$:
%
\; ,
%
\begin{eqnarray}
I(T,T)_{(\lambda_{V} = \lambda_{\gamma})} &=& m_{q}^{2} \Phi_{2} + \Big [z^{2}+(1-z)^{2} \Big] (  \bk \bPhi_{1})
\\
\nonumber
&& + \frac {m_{q}}{M+2m_{q}} \Big [ \bk^{2}  \Phi_{2}-(2z-1)^{2}(\bk \bPhi_{1})  \Big ]
\; .
\end{eqnarray}

\item the helicity flip by one unit, i.e. from the transverse photon $\lambda_\gamma = \pm 1$ to the longitudinally polarized meson, $\lambda_V = 0$.
\begin{eqnarray}
I(L,T) &=& -2Mz(1-z)(2z-1)(\be \bPhi_{1}) \Big [1+ \frac{(1-2z)^{2}}{4z(1-z)}\frac{2m_{q}}{M+2m_{q}} \Big ] 
\\
\nonumber
&& + \frac {Mm_{q}}{M+2m_{q}} (2z-1) (\be \bk) \Phi_{2}
\; .
\end{eqnarray}
\item the helicity flip by two units, from the transverse photon $\lambda_\gamma = \pm1$ to the transversely polarized meson with $\lambda_V = \mp 1$:
\begin{eqnarray}
I(T,T)_{(\lambda_{V} = - \lambda_{\gamma})} &=& 2z (1-z)(\Phi_{1x} k_{x}  - \Phi_{1y} k_{y} ) \nonumber
\\ 
&& - \frac {m_{q}}{M+2m_{q}} \Big [(k_{x}^{2} -k_{y}^{2}) \Phi_{2}-(2z-1)^{2}(k_{x}\Phi_{1x} -k_{y}\Phi_{1y})  \Big ]
\; .
\end{eqnarray}
\end{enumerate}
Here 
\begin{eqnarray}
M = \sqrt{\bk^2 + m_q^2 \over z (1-z)} \, 
\end{eqnarray}
is the invariant mass of the $q \bar q$ pair.
In the expressions above $\bPhi_{1}$, $\Phi_{2}$ are given by (see e.g. Ref. \cite{Ivanov:2004ax}):
\begin{eqnarray}
\Phi_{2} &=& -{1
\over (\br+\bkappa)^2 + \varepsilon^2} -{1 \over
(\br-\bkappa)^2 + \varepsilon^2} + {1 \over (\br +
\bDelta/2)^2 + \varepsilon^2} + {1 \over (\br -
\bDelta/2)^2 + \varepsilon^2} \, ,
\nonumber \\
\bPhi_{1} &=&
-{\br + \bkappa \over (\br+\bkappa)^2 +
\varepsilon^2} -{\br - \bkappa \over (\br-\bkappa)^2
+ \varepsilon^2} + {\br + \bDelta/2 \over (\br +
\bDelta/2)^2 + \varepsilon^2} + {\br - \bDelta/2 \over
(\br - \bDelta/2)^2 + \varepsilon^2} \, .
\end{eqnarray}
Quark and antiquark carry a fraction $z$ and $1-z$ of the vector meson lightcone plus-momentum as well as the  transverse momenta $\bk_q = \bk + z \bDelta$ and $\bk_{\bar q} = -\bk + (1-z) \bDelta$. Furthermore, we introduced 
$\br = (\bk_q - \bk_{\bar q})/2 = \bk + (z - \frac{1}{2}) \bDelta$,
and $\varepsilon^{2} = 
m_{q}^{2} + z (1 - z) Q^{2}$.
The $\rho$--meson is treated as the pure $s$-wave bound state of light quarks with the constituent quark mass taken as $m_q = 0.22 \, \rm{GeV}$.
As to the vector meson radial light-front wave function (LFWF), we use the parametrization 
\begin{eqnarray}
\psi_V(z,k) = C \exp \Big[ - \half a \, \vec k^2  \Big] ,
\end{eqnarray}
with
\begin{eqnarray}
\vec k^2 ={1 \over 4} (M^2 - 4 m_q^2) \, .
\end{eqnarray}
The wave function parameter $a$ is adjusted such that the leptonic decay width of the $\rho$ meson is reproduced.

We now wish to specify the function ${\cal F}(x, \bkappa_1,\bkappa_2)$. Here the transverse momenta sum up to the total transverse momentum transfer $\bkappa_1 + \bkappa_2 = \bDelta$. We parametrize them as
\begin{eqnarray}
\bkappa_1 = {\bDelta \over 2} +  \bkappa \, , \, \bkappa_2 = {\bDelta \over 2} -  \bkappa \, ,
\end{eqnarray}
in terms of the integration variable $\bkappa$.
Then, we write 
\begin{eqnarray}
{\cal F} \Big(x, {\bDelta \over 2} + \bkappa, {\bDelta \over 2} - \bkappa \Big ) = f(x,\bkappa) \, G(\bDelta^2) \, ,
\label{eq:off-diagonal-glue}
\end{eqnarray}
where $f(x,\bkappa)$ is an unintegrated gluon distribution, which in the perturbative domain at large $\bkappa^2$ is related to the standard gluon distribution as
\begin{eqnarray}
f(x,\bkappa) \to {\partial x g(x,\bkappa^2) \over \partial \log \bkappa^2} \, .
\end{eqnarray}
For the function $G(\bDelta^2)$, which satisfies $G(0) =1$ and takes into account the momentum transfer dependence of the proton-Pomeron coupling, we adopt two options:
\begin{enumerate}
    \item an exponential parametrization: \begin{eqnarray}
    G(\bDelta^2) = \exp\Big[ - \half B \bDelta^2 \Big] \, ,
    \end{eqnarray}
    with a diffraction slope of $B = 4 \, \rm{GeV}^{-2}$.
    \item a dipole form factor parametrization often used in nonperturbative Pomeron models \cite{Donnachie:2002en}:
    \begin{eqnarray}
    G(\bDelta^2) = {4 m_p^2 + 2.79 \bDelta^2 \over 4 m_p^2 + \bDelta^2} \, {1 \over \Big( 1 + {\bDelta^2 \over \Lambda^2} \Big)^2} \, ,
    \end{eqnarray}
    with $\Lambda^2 = 0.71 \, \rm{GeV}^2$.
\end{enumerate}
Besides the model of Eq.(~\ref{eq:off-diagonal-glue}), we also tried a parametrization of the off-diagonal UGD of the type
\begin{eqnarray}
{\cal F}(x, \bkappa_1,\bkappa_2) = 
{2 \bkappa_1^2 \bkappa_2^2 \over \bkappa_1^4 + \bkappa_2^4} \, f\Big(x,{\bkappa_1^2 + \bkappa_2^2 \over 2}\Big) \, G(\bDelta^2) \, ,
\end{eqnarray}
which had been proposed in Ref.~\cite{Cudell:2009wck}. As we found only small differences, we stick to the simpler form of Eq.(~\ref{eq:off-diagonal-glue}).

Before we come to a discussion of our results for the differential cross section, let us briefly discuss the 
different models for the unintegrated gluon distribution which we applied. Once again we stress that our process is dominated by the nonperturbative domain. 
Here our assumption is simply that this effective two-gluon exchange
couples to the $q \bar q$ pair conserving helicities of quarks. This rather economical choice avoids the introduction of additional parameters. 
Let us remind the reader that also a class of nonperturbative models can be cast into the form based on a nonperturbative gluon distribution \cite{Shoshi:2002fq}.
Other models including an anomalous chromomagnetic moment of quarks or other couplings motivated by an instanton model of the QCD vacuum  can be found for example in \cite{Korchagin:2011wy,Kochelev:2015pqd}. 
\begin{figure}[h]
\centering
\includegraphics[width=.45\textwidth]{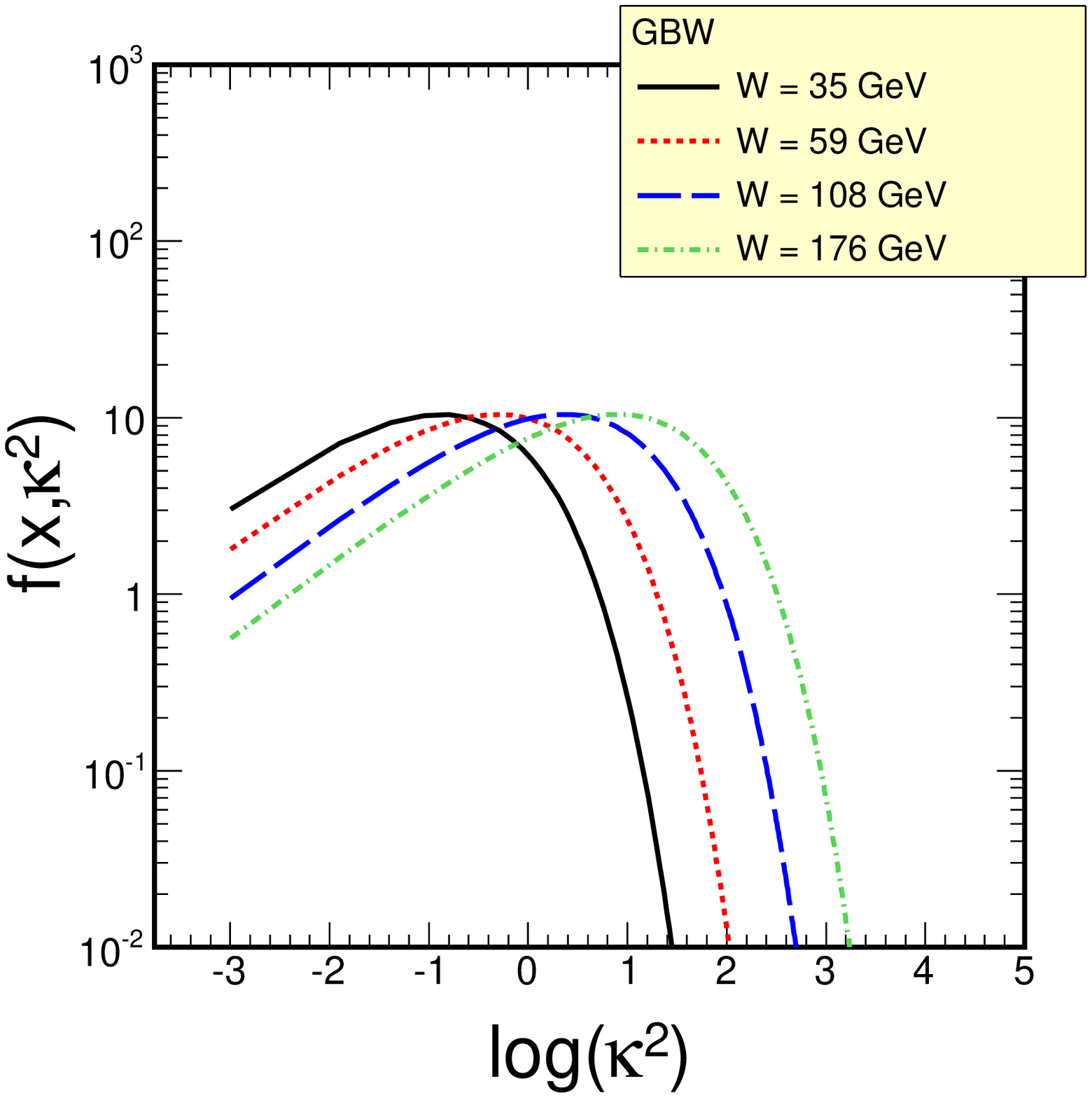}
\includegraphics[width=.45\textwidth]{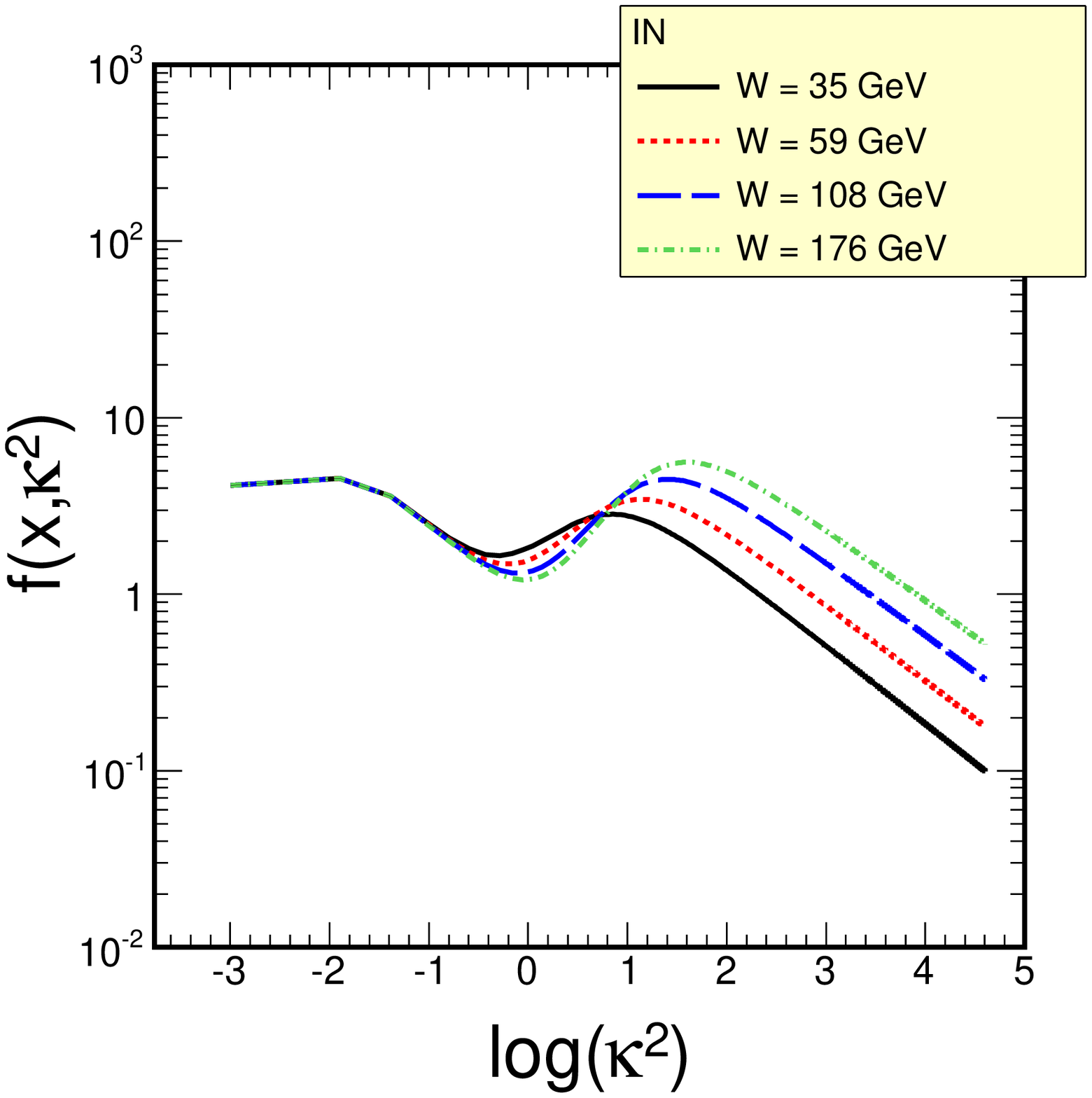}
\includegraphics[width=.45\textwidth]{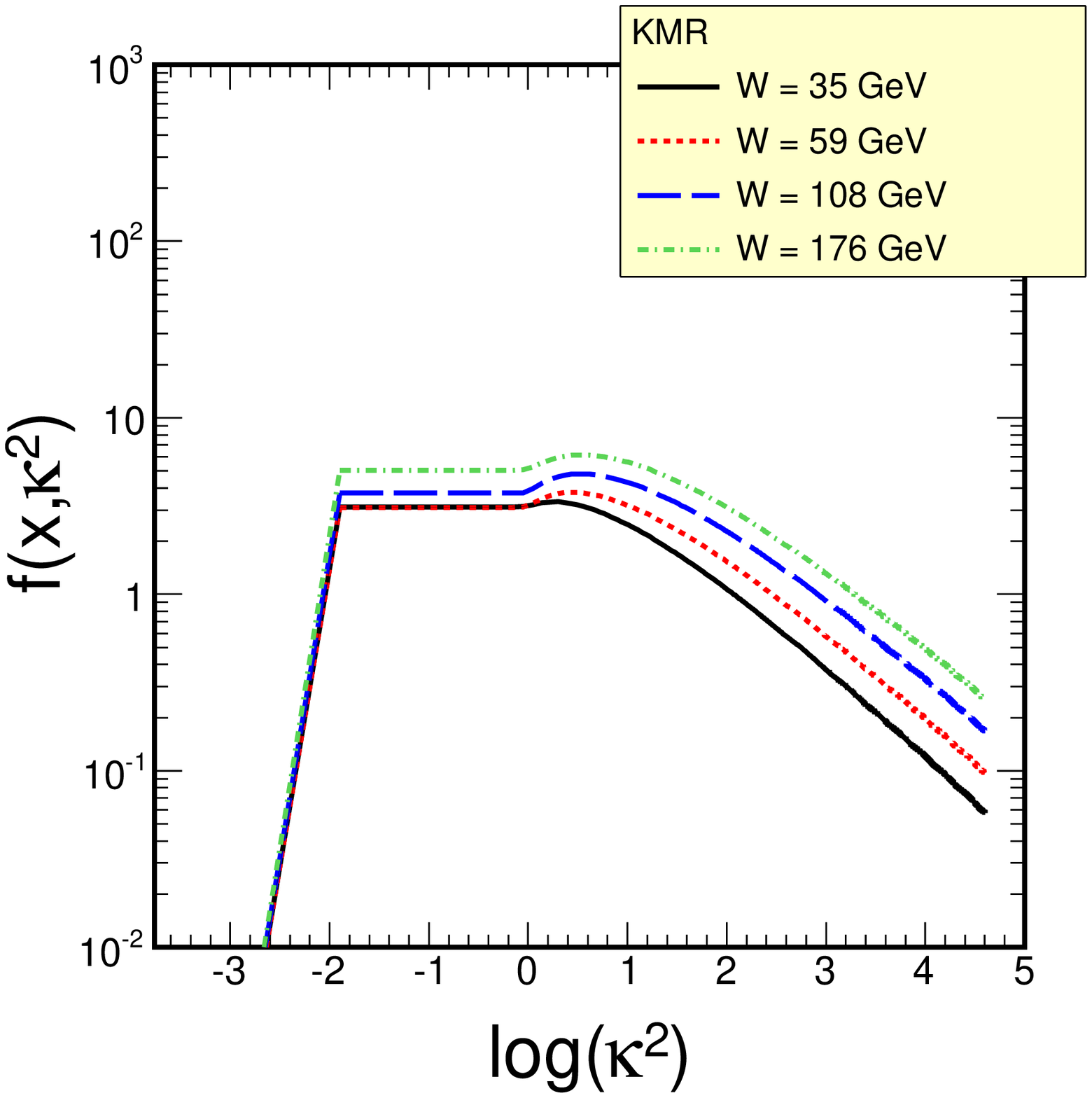}
\includegraphics[width=.45\textwidth]{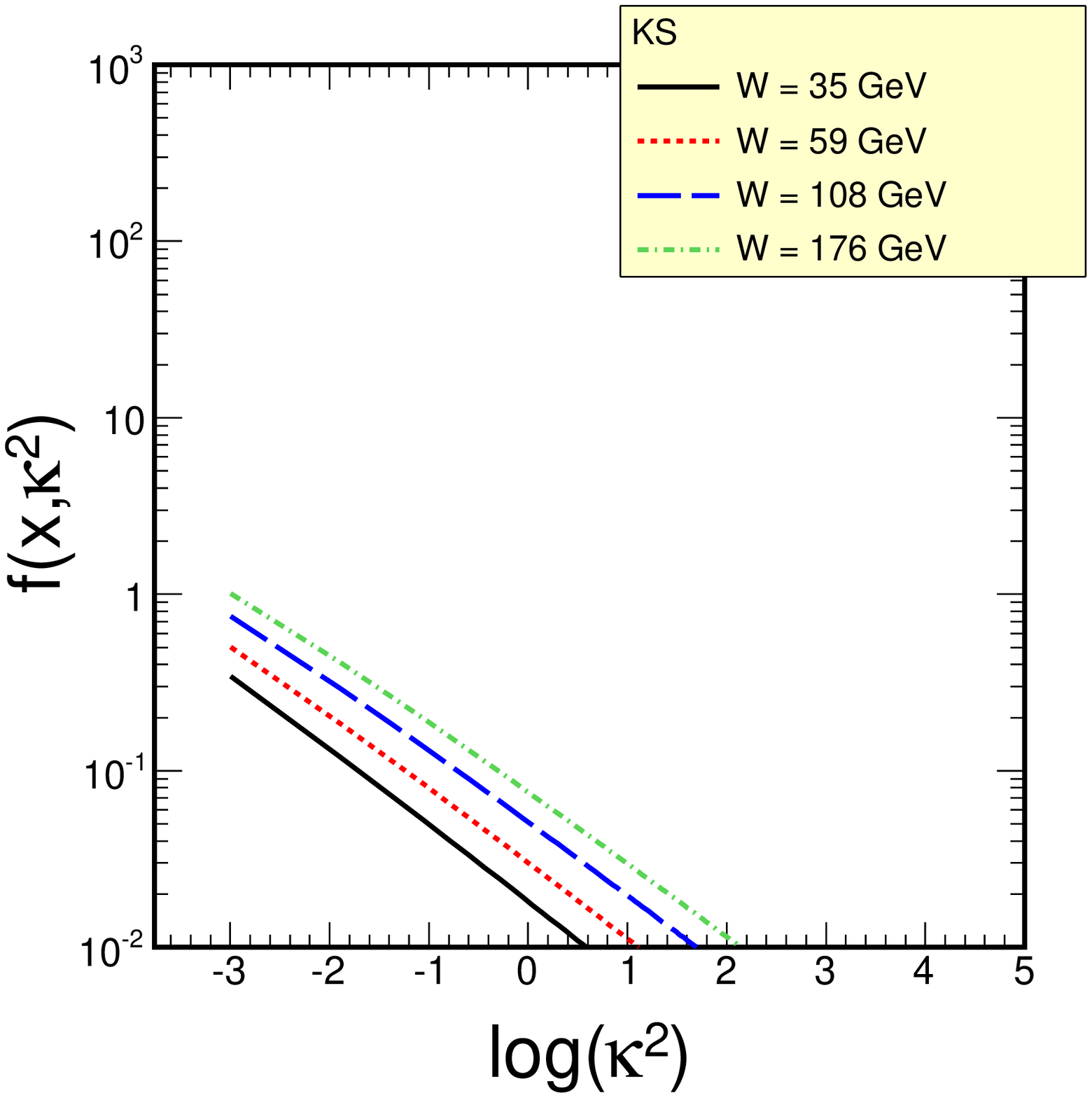}
\caption{Different models of UGDs in $\log \kappa^{2}$ distributions.
The results for different x (or energy) are shown.}
\label{fig:UGDs}
\end{figure}
We will perform our numerical calculations for a number of UGDs. 
Those will be:
\begin{enumerate}
    \item a UGD obtained from the Fourier transform of the Golec-Biernat--W\"usthoff dipole cross section~\cite{Golec-Biernat:1998zce}. We slightly change the normalization and use fixed value of $\alpha_s = 0.3$. 
    \item a UGD by Ivanov and Nikolaev~\cite{Ivanov:2000cm}, which contains explicit modeling of the nonperturbative low $\kappa^2$ part. Here, as for all the UGDs below, a freezing of $\alpha_s$ in the soft region is employed.
    \item a UGD of the Durham group constructed by the so-called Kimber-Martin-Ryskin (KMR)~\cite{Kimber:2001sc} procedure starting from collinear partons of Ref.~\cite{Harland-Lang:2014zoa}. The UGD is extended to the soft region in a simple way, such that the distribution in $\log \kappa^2$ levels off to an $x$-dependent plateau at small $\kappa^2$. 
    \item The Kutak-Sta\'sto (KS) nonlinear UGD from Ref.~\cite{Kutak:2004ym}.
\end{enumerate}
In Fig.~\ref{fig:UGDs} we show the four different versions of the UGDs which we use for the $x$-values corresponding to the $\gamma p$-energies relevant to the CMS data \cite{CMS:2018ixo}.

\section{Results}

Most phenomenological studies of vector meson photoproduction concentrate on the modelling of the forward amplitude or on the region of small momentum transfers within the diffraction cone. 
The CMS data for $d \sigma/dt$ up to $-t \sim 1 \, \rm{GeV}^2$ present us with welcome opportunity to test popular models of vector meson production in a broader kinematic region.

It should be noted, that we are still far away from the regime of the ``hard exclusive'' process for which $s \sim |t| \sim |u| \gg \Lambda_{QCD}^2$, and where a perturbative approach based on a different type of factorization becomes applicable \cite{Lepage:1980fj}. Rather, although we go beyond the diffraction cone, we are still in the domain of Regge asymptotics, $-t/s \ll 1$. 

We come to the comparison with the results of the CMS collaboration \cite{CMS:2018ixo}. The differential cross section $d \sigma / dt$ was extracted from ultraperipheral proton-lead collisions for four different $\gamma p$ cm-energies: $W=35, 59, 108 ,179 \, \rm{GeV}$.
At these energies we can safely approximate $t \sim - \bDelta^2$.

\begin{figure}[h]
\centering
\includegraphics[width=8.cm]{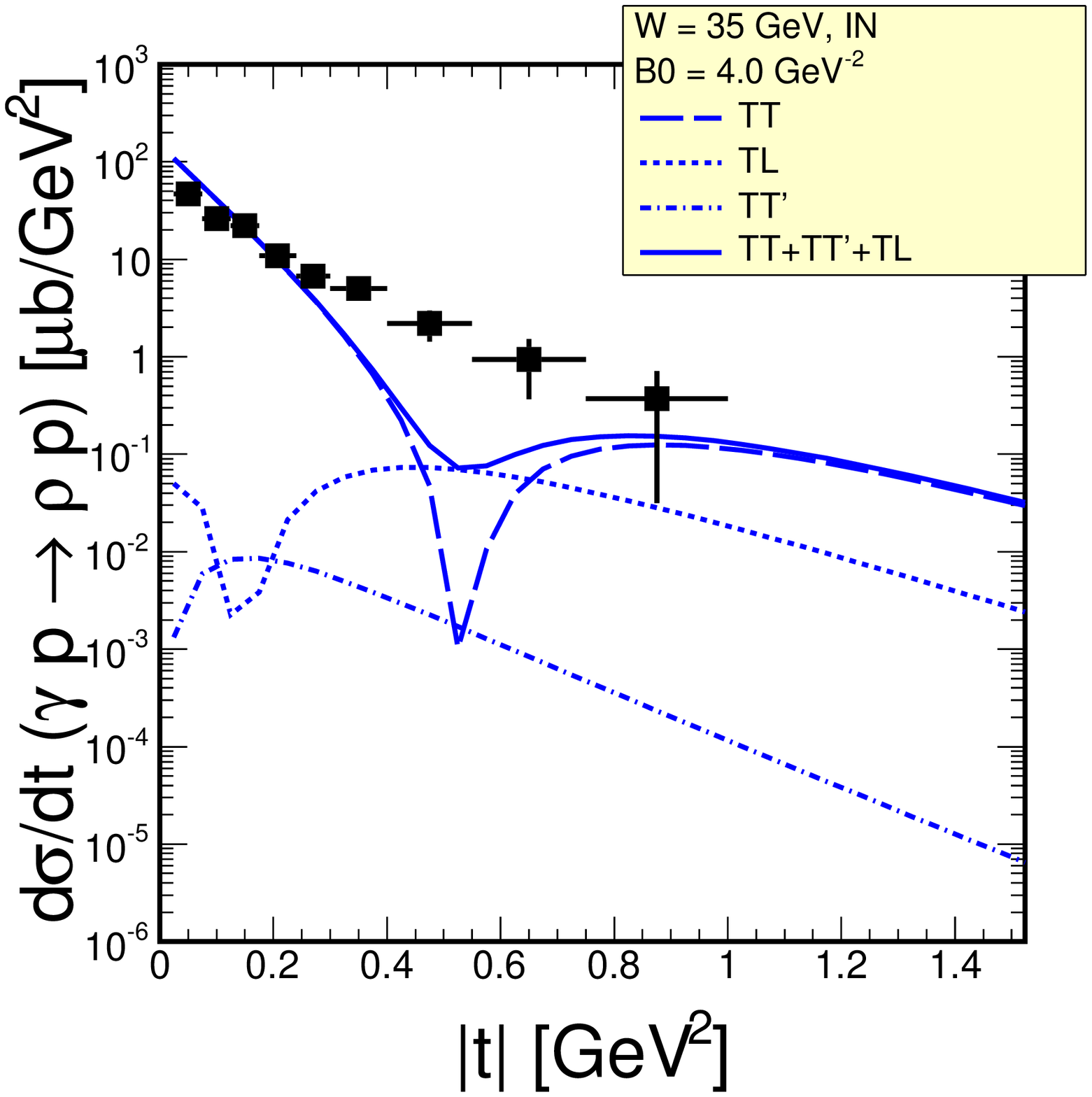}
\includegraphics[width=8.cm]{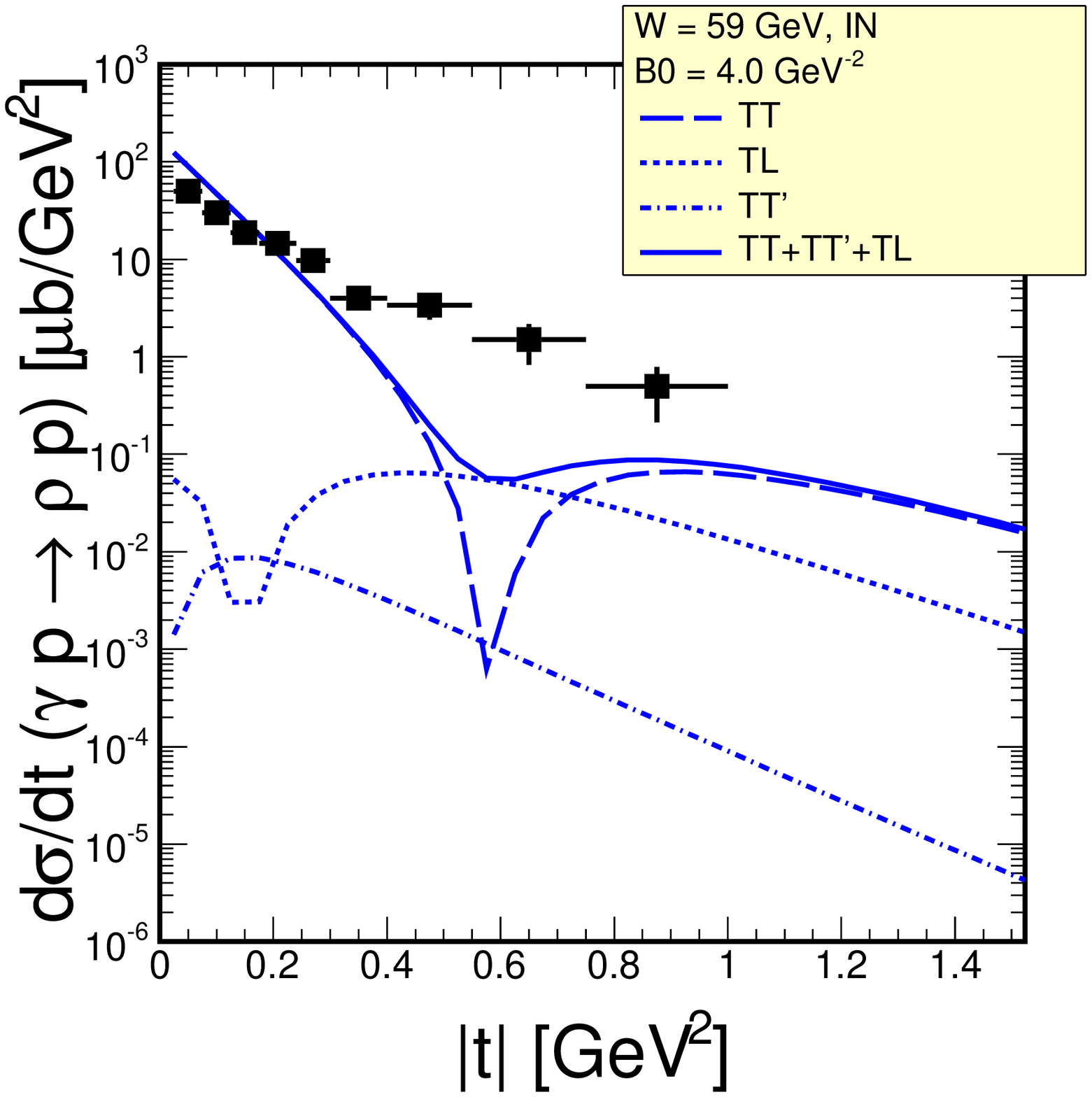}
\includegraphics[width=8.cm]{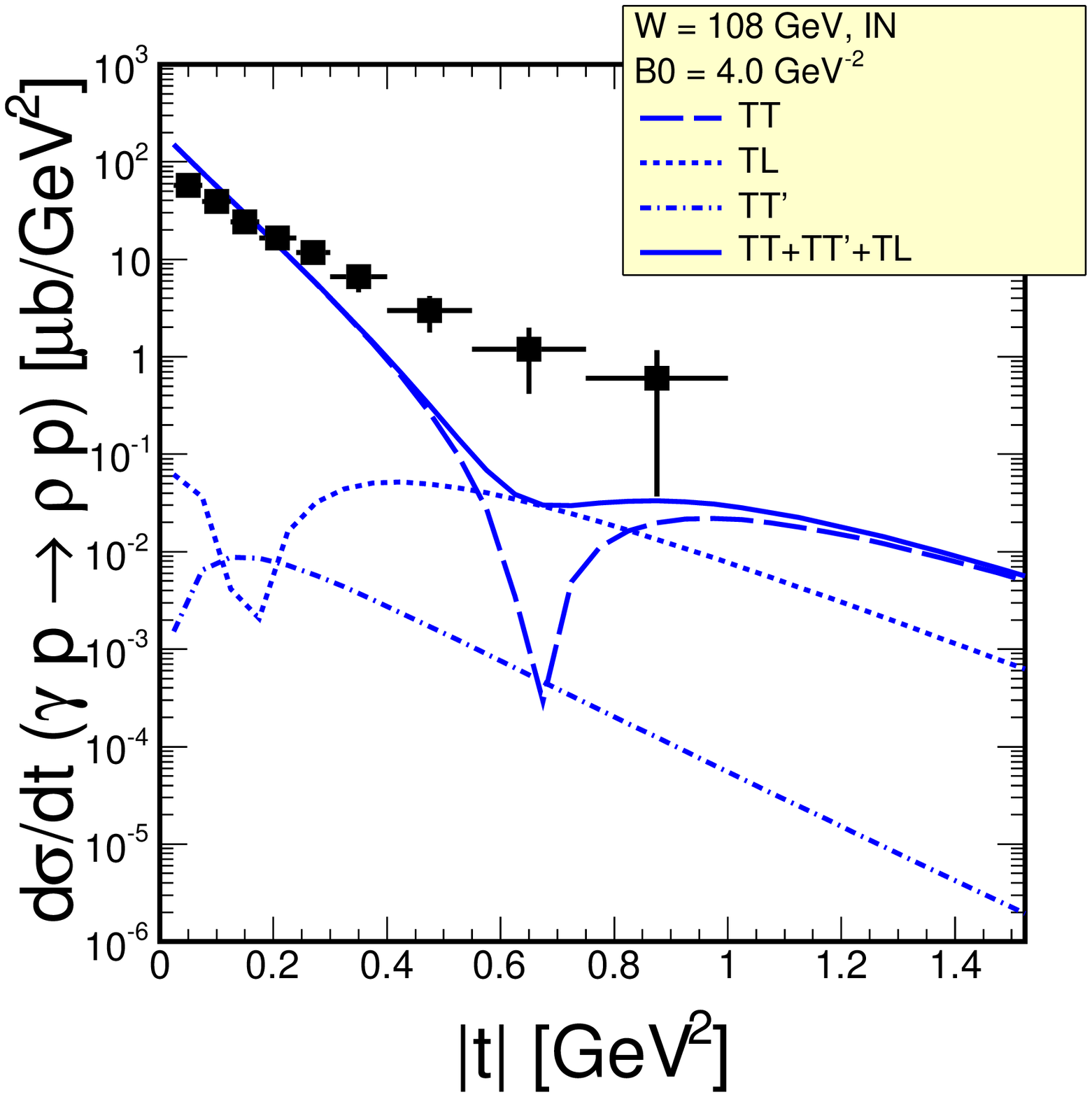}
\includegraphics[width=8.cm]{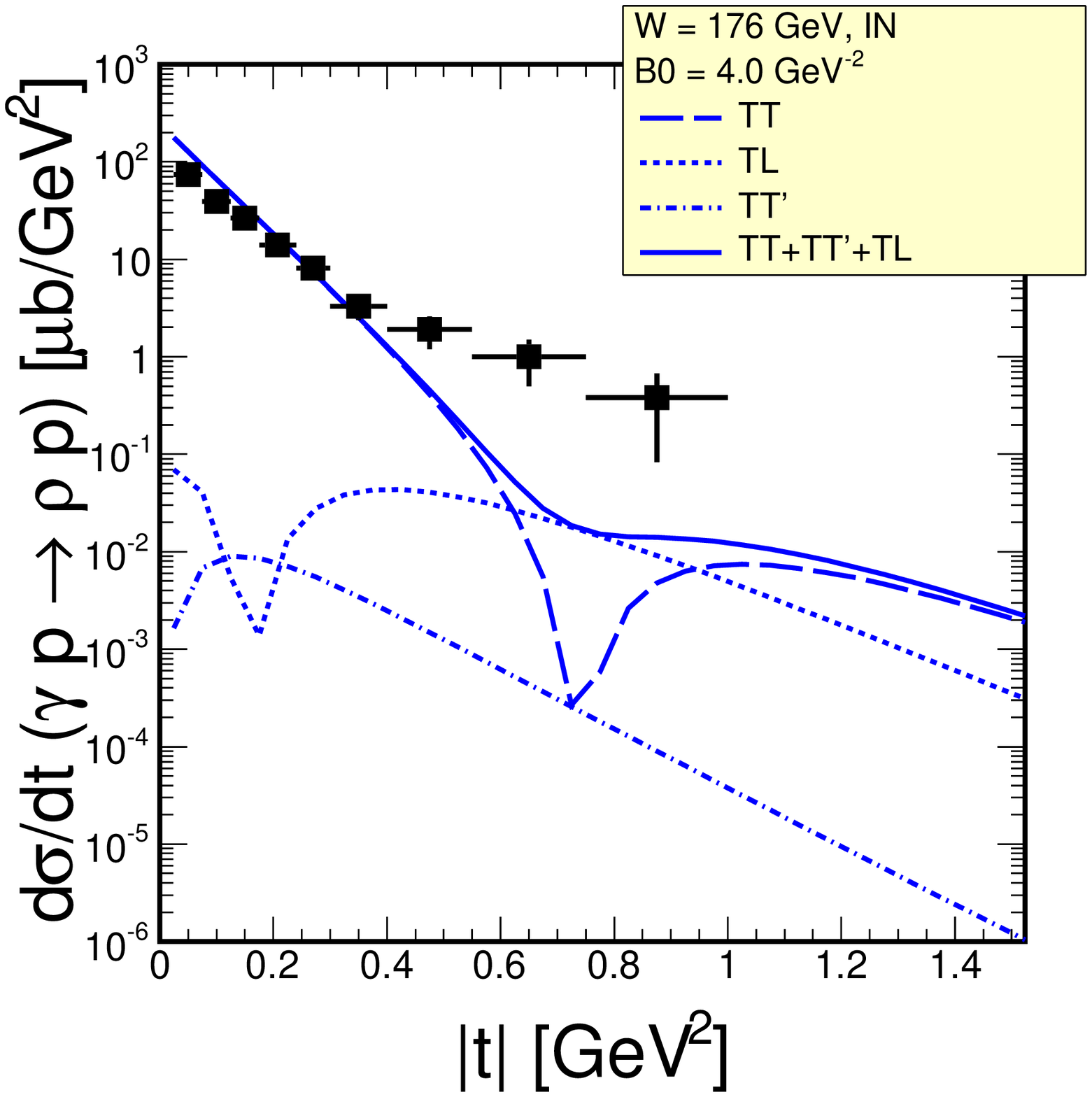}
\caption{Distribution in $t$, the four-momentum transfer squared in the
$\gamma p \to \rho p$ reaction for different energies and the
Ivanov-Nikolaev UGD. Here the exponential parametrization of the form factor $G(\bDelta^2)$ was used.
}
\label{fig:IN_exponential}
\end{figure}

\begin{figure}[h]
\centering
\includegraphics[width=8.cm]{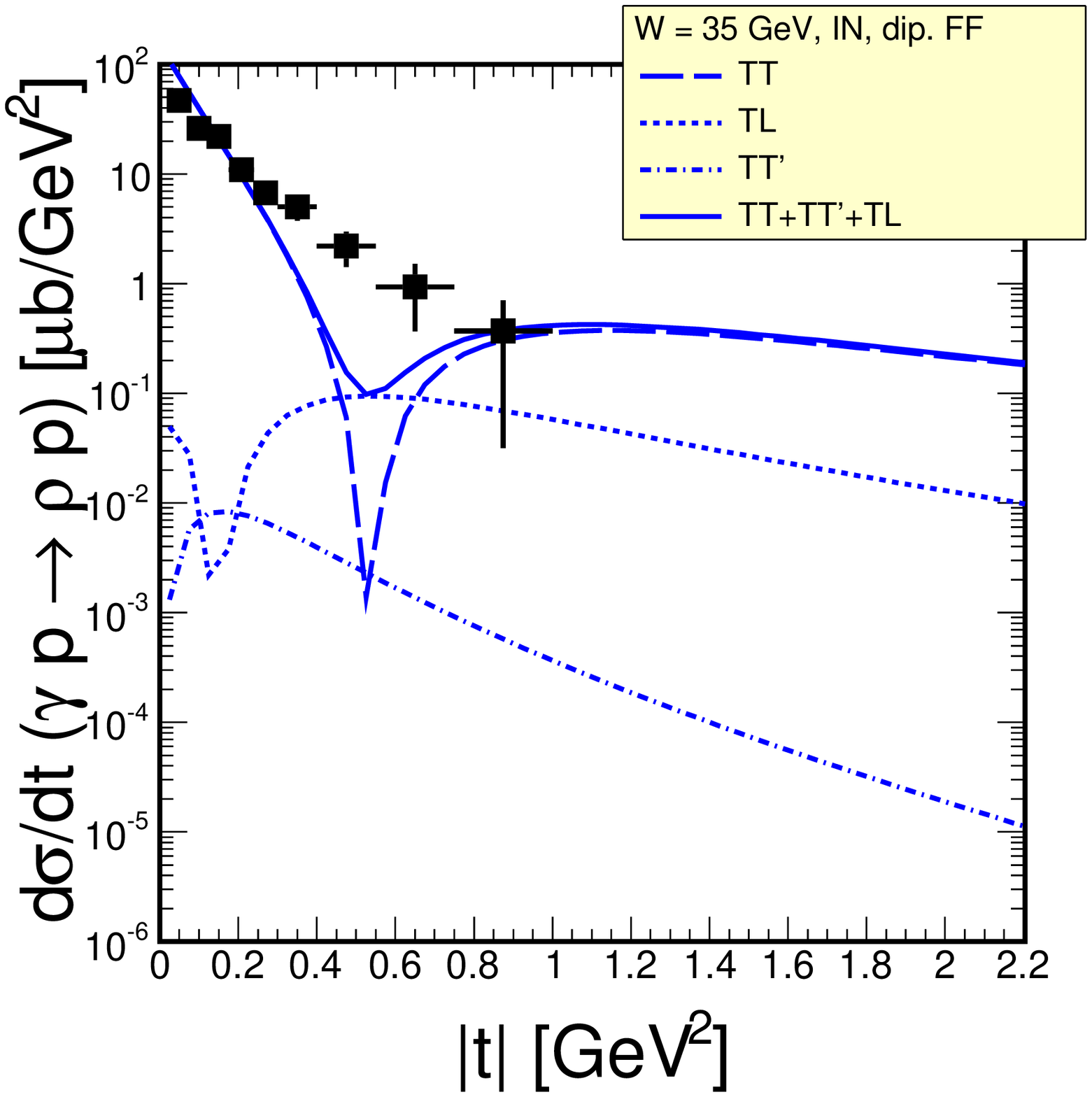}
\includegraphics[width=8.cm]{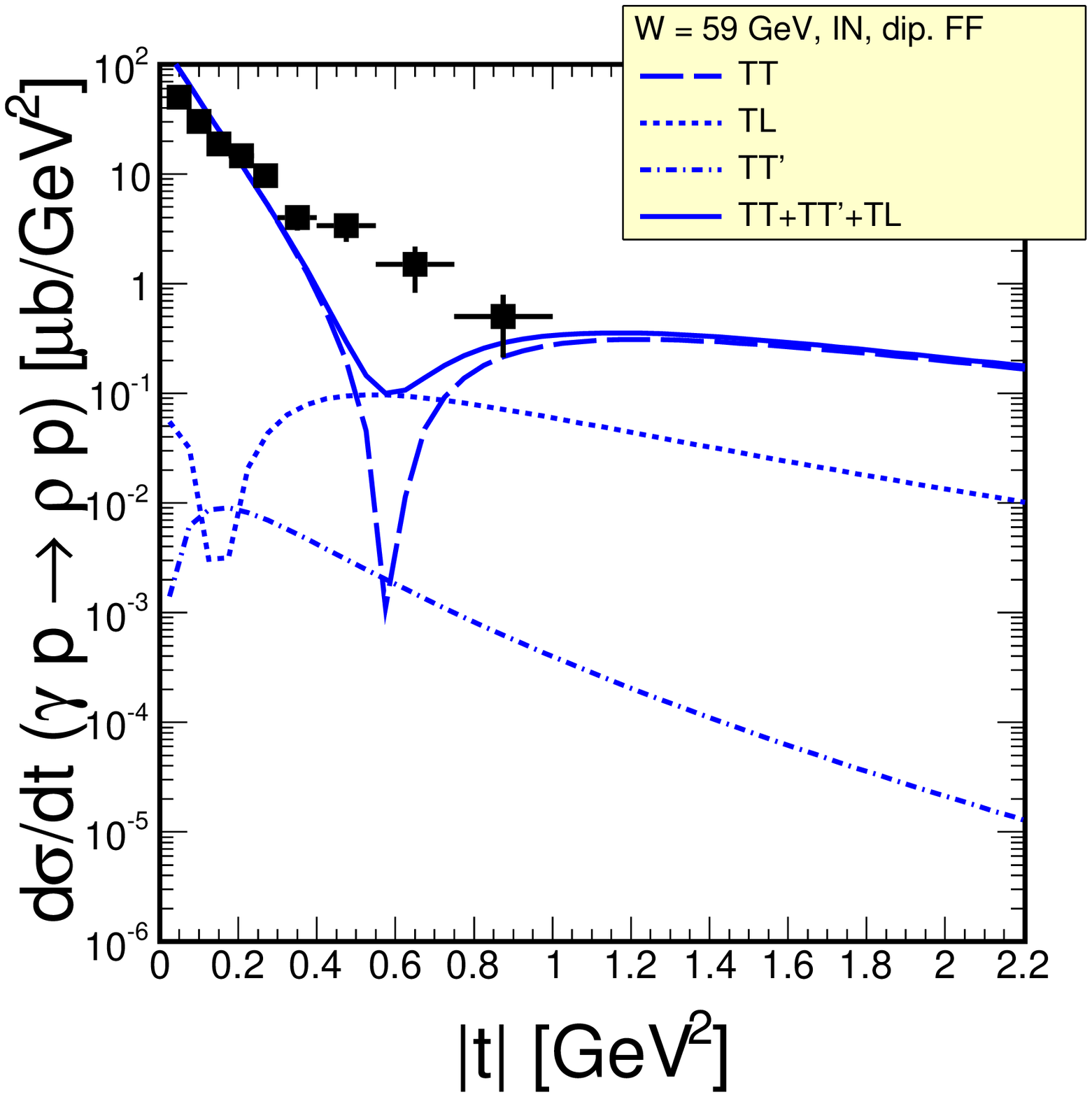}
\includegraphics[width=8.cm]{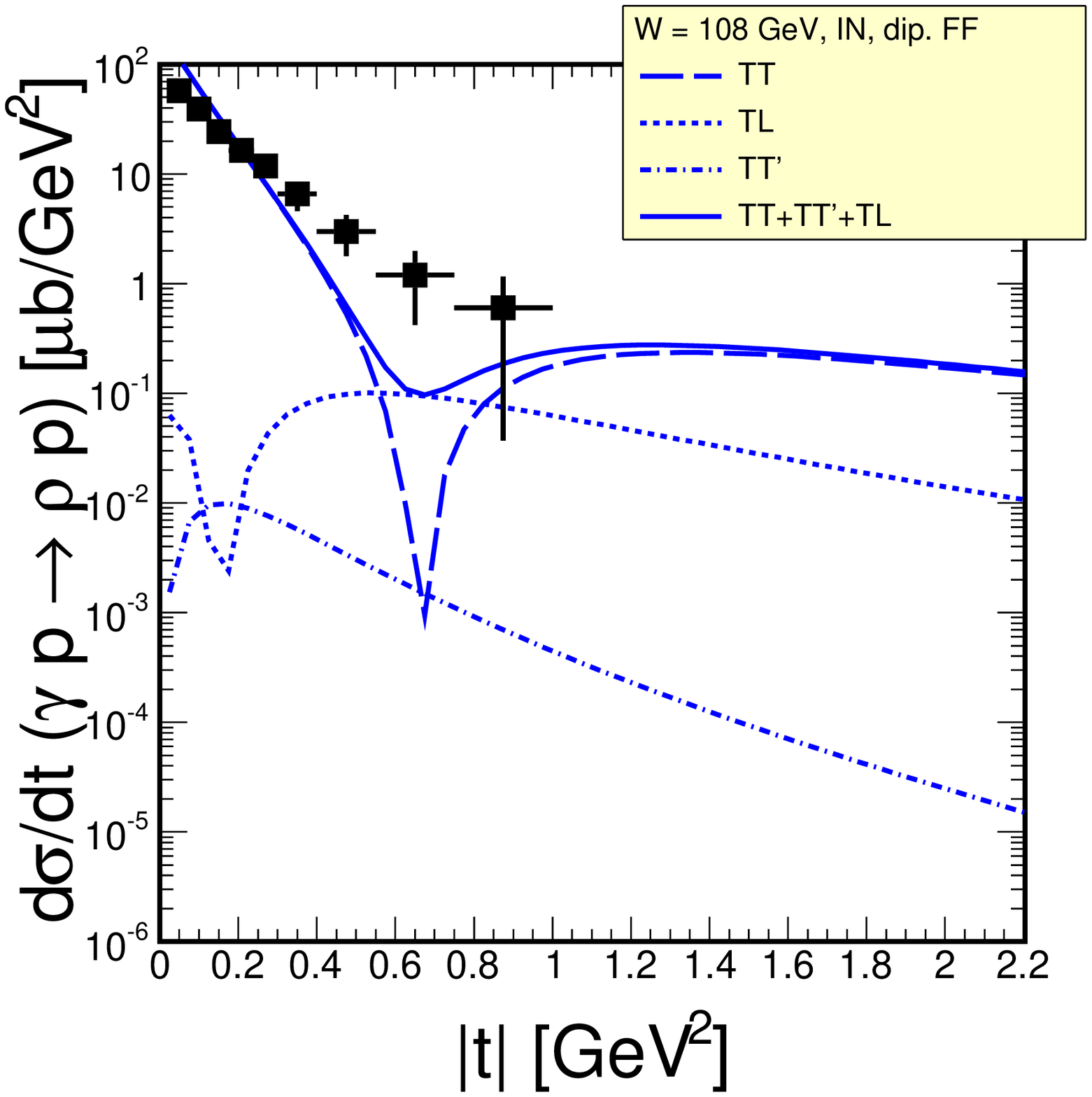}
\includegraphics[width=8.cm]{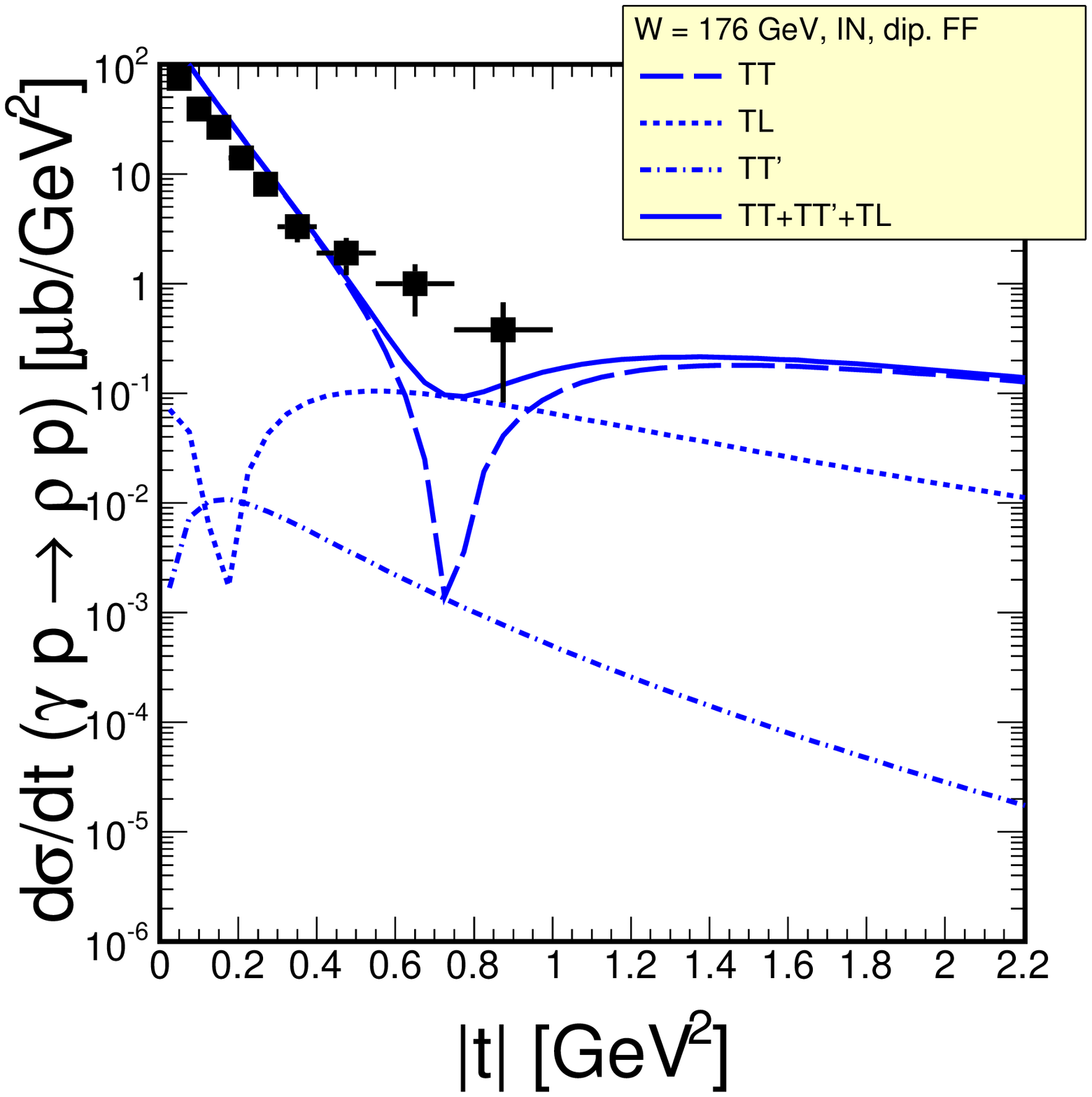}
\caption{Distribution in $t$, the four-momentum transfer squared in the
$\gamma p \to \rho p$ reaction for different energies and the
Ivanov-Nikolaev UGDF. Here the dipole parametrization of the form factor $G(\bDelta^2)$ was used. }
\label{fig:IN_dipole}
\end{figure}

In Fig.~\ref{fig:IN_exponential}, we show our results
for the IN UGD using the exponential parametrization for $G(\bDelta^2)$. We show the $T \to T$ helicity conserving contribution by the long dashed line. The dotted line shows the $T \to L$ transition, where helicity is changed by one unit. Finally, by the dash-dotted line we display the $T \to T'$ transition where $|\lambda_\gamma - \lambda_V| = 2$.
We stress, that we take the Pomeron coupling on the proton side to be helicity conserving, in accord with 
the very small spin-flip coupling found in Regge analyses \cite{Buttimore:1998rj}. In this case, in agreement with general requirements of angular momentum conservation, helicity flip amplitudes vanish in the forward direction $\propto |\bDelta|^{|\lambda_\gamma - \lambda_V|}$. Notice that this vanishing of the spin flip contributions is not visible at the resolution of the plot in Fig.~\ref{fig:IN_dipole}.

We observe, that the $T \to T$ contribution has a dip at $-t \sim 0.5 \div 0.7 \, \rm{GeV}^{-2}$, which position is slightly dependent on energy. It derives from a zero in the corresponding amplitude. 
We note that the dip is partially filled by the $T \to L$ contribution and turns into a shoulder with increasing energy. The $T \to L$ transition itself possesses a dip at $-t \lsim 0.2 \, \rm{GeV}^{-2}$ but in this region the $T \to T$ transition vastly dominates. The double helicity flip contribution is very small throughout the whole kinematic region.

In Fig.~\ref{fig:IN_dipole} we show the results for the same IN UGD, but now using the dipole parametrization for $G(\bDelta^2)$. We observe that some qualitative features, like the position of the dip are very similar to the previous case, however the cross section develops a much harder tail at large $-t$. In general, the IN UGD gives a reasonable description of data, especially within the diffraction cone, however a diffractive minimum or shoulder predicted by this glue are not borne out by the data.

We now turn to the results for other UGDs, where for definiteness we stick to the dipole parametrisation of $G(\bDelta^2)$.

\begin{figure}[h]
\centering
\includegraphics[width=8.cm]{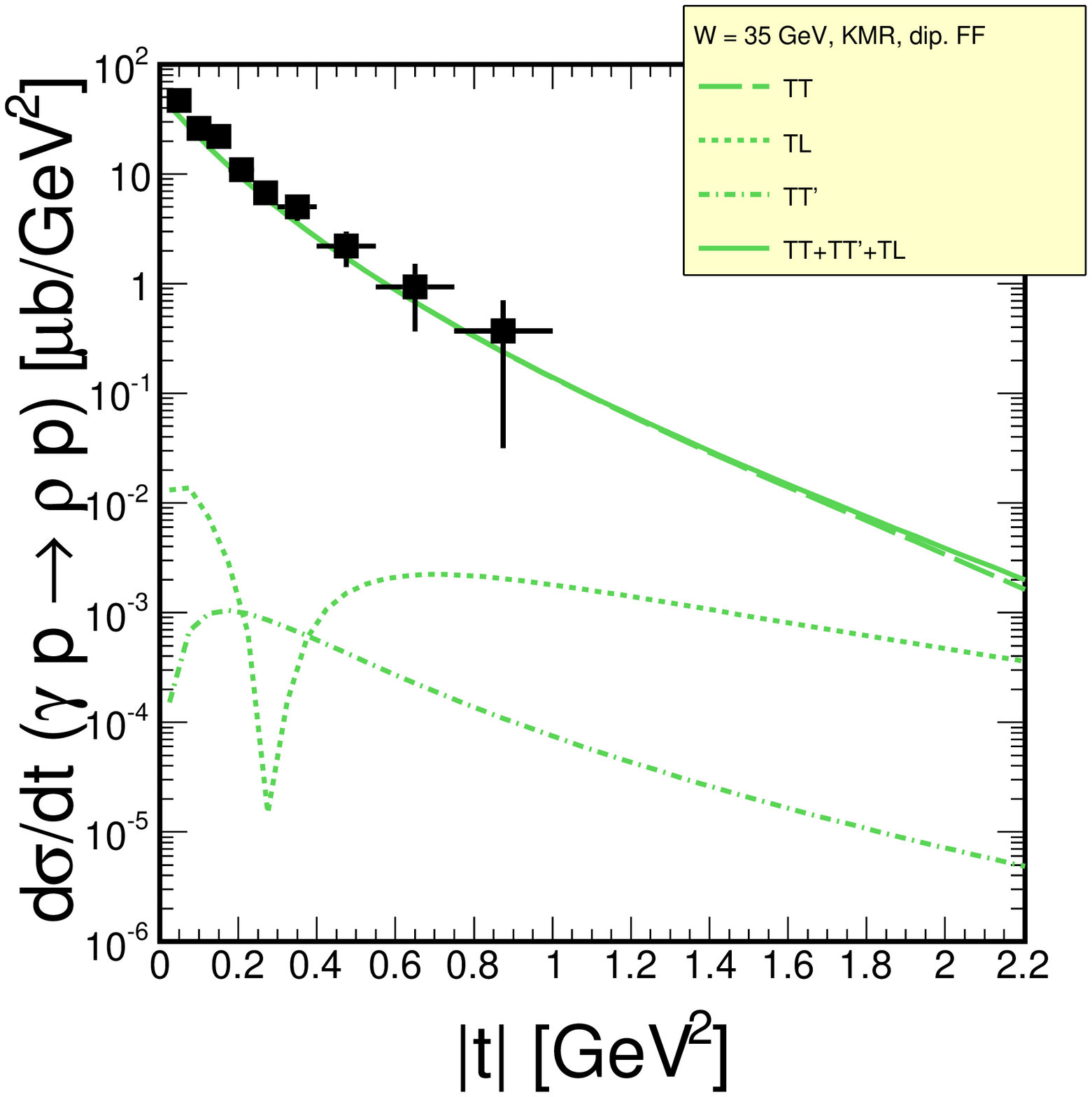}
\includegraphics[width=8.cm]{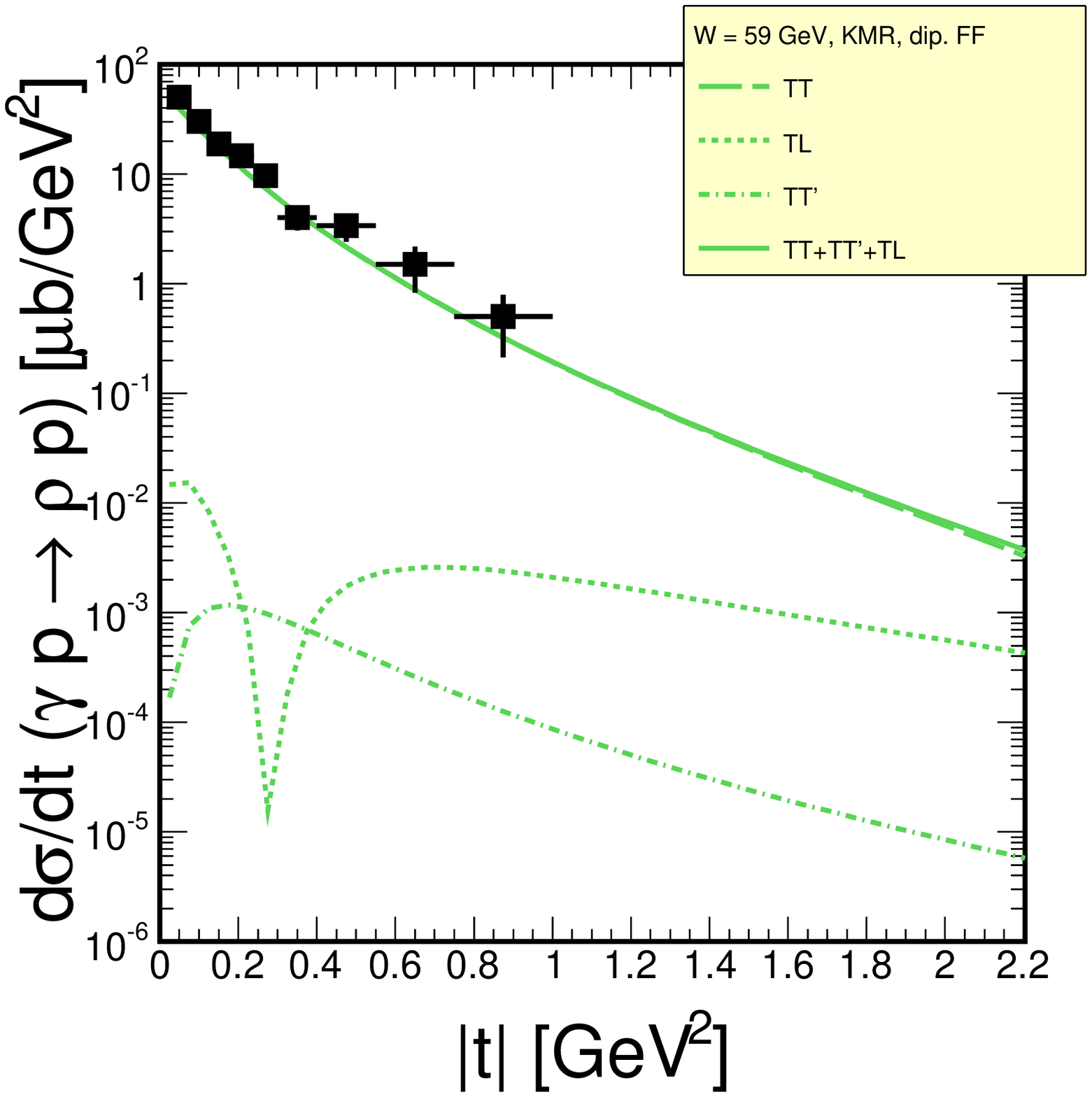}
\includegraphics[width=8.cm]{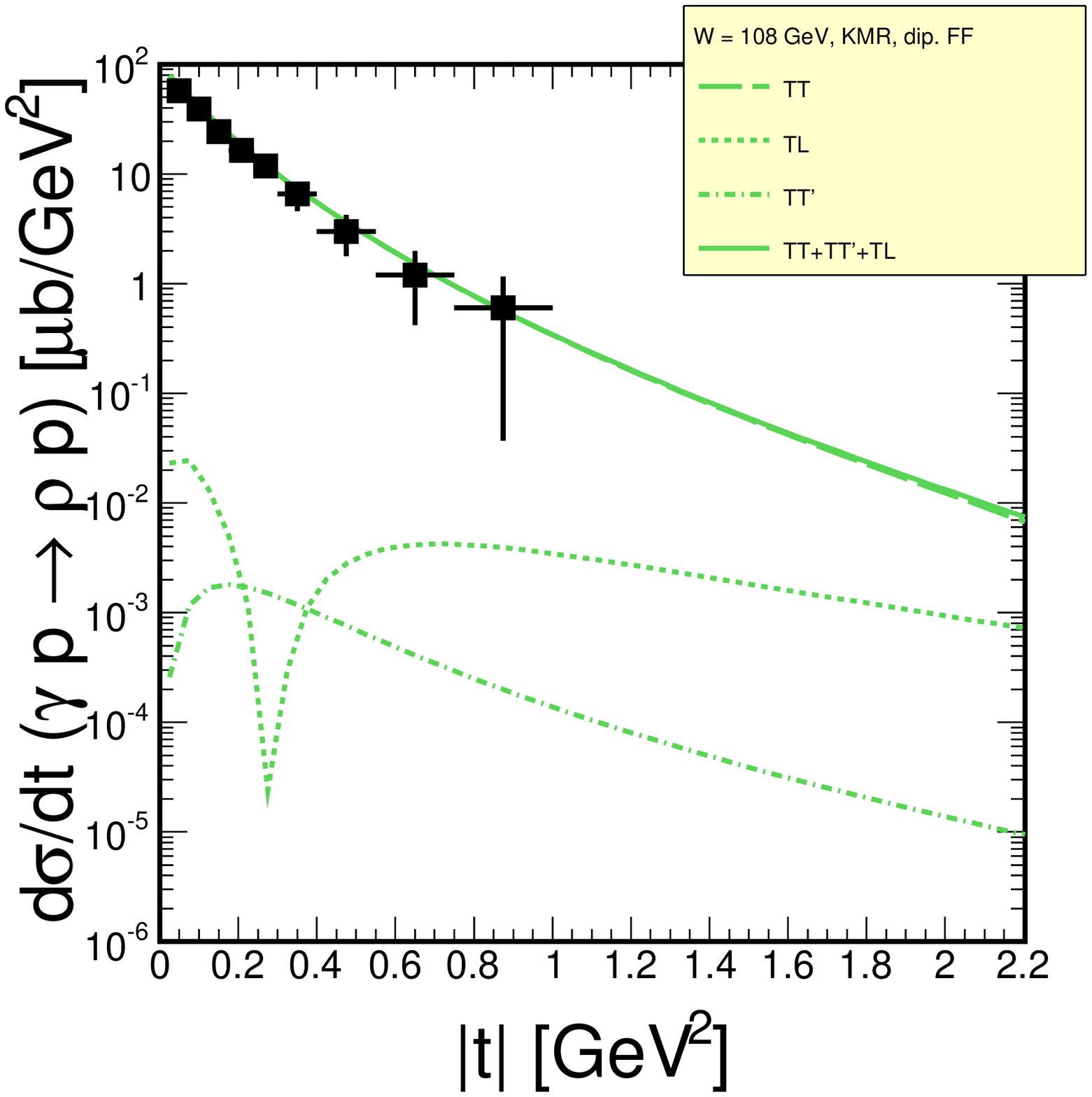}
\includegraphics[width=8.cm]{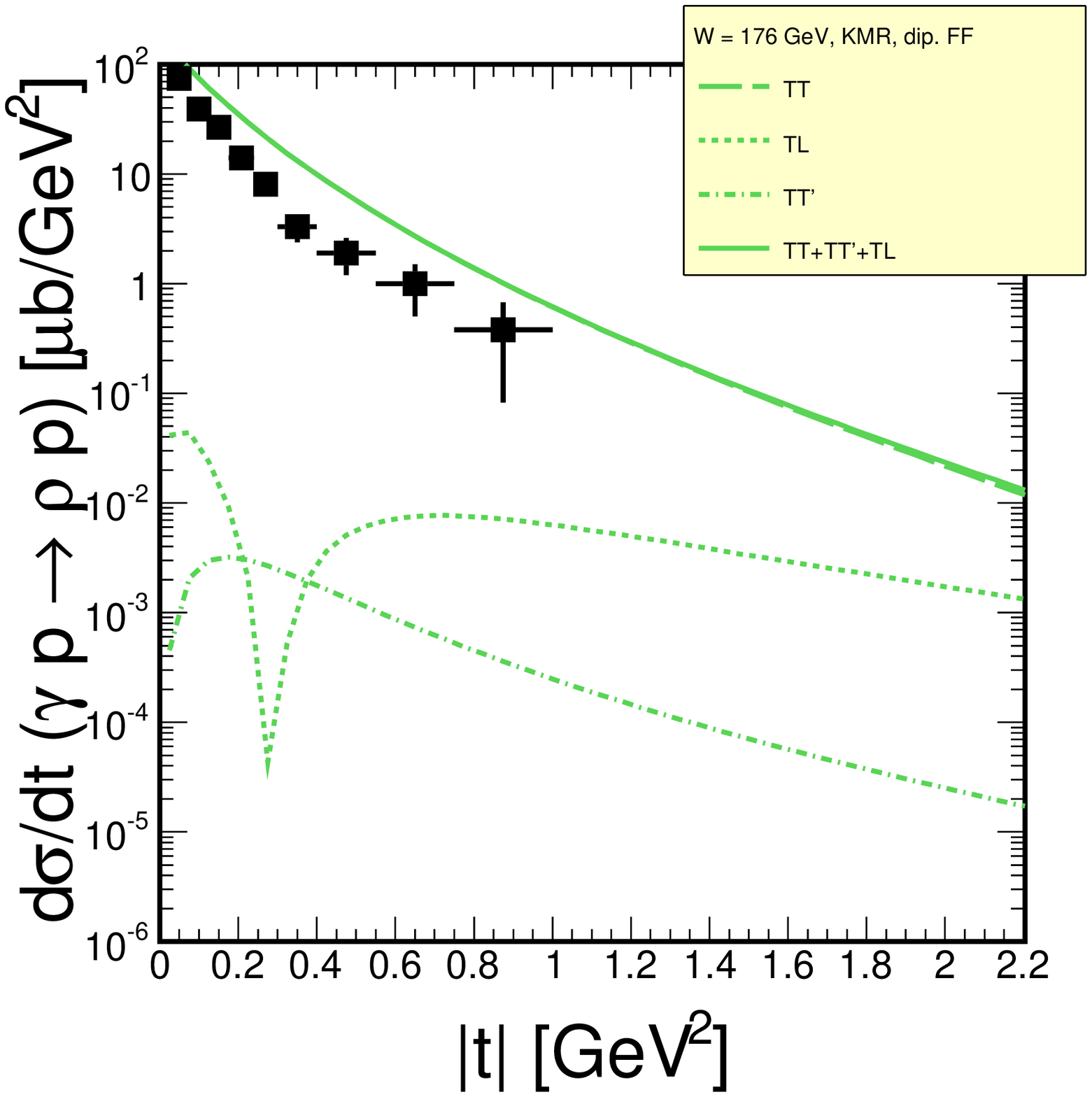}
\caption{Distribution in $t$ for different energies and the KMR
  UGD. Here the dipole parametrization of the form factor 
$G(\bDelta^2)$ was used.
}
\label{fig:dsig_KMR}
\end{figure}

In Fig.~\ref{fig:dsig_KMR} we show the results for the KMR UGD. Here the helicity conserving part dominates throughout. In this case, there is no dip in the differential cross section.
The description of data is very good except for the highest energy, where the $t$-dependence is too hard.

\begin{figure}[h]
\centering
\includegraphics[width=8.cm]{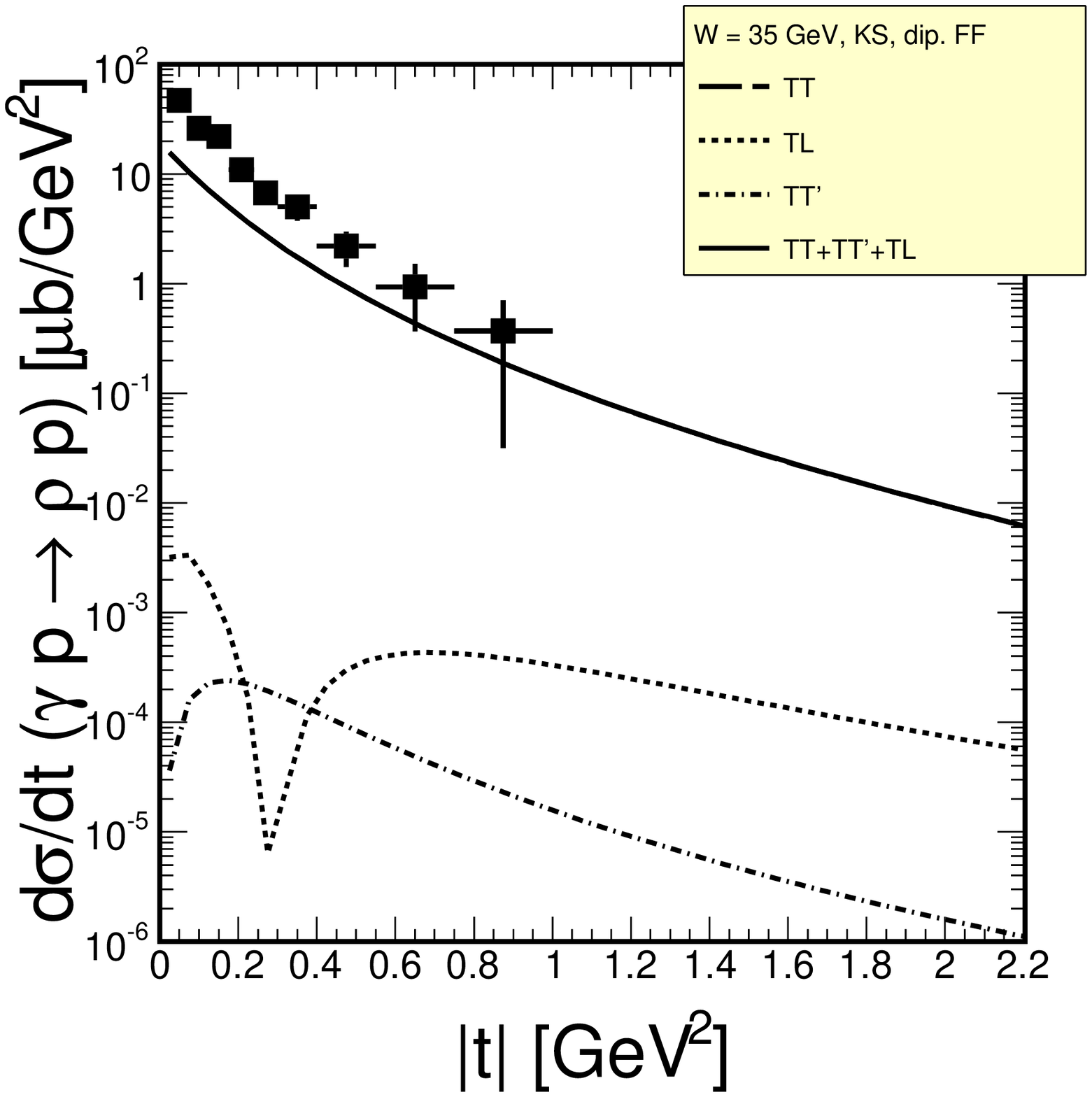}
\includegraphics[width=8.cm]{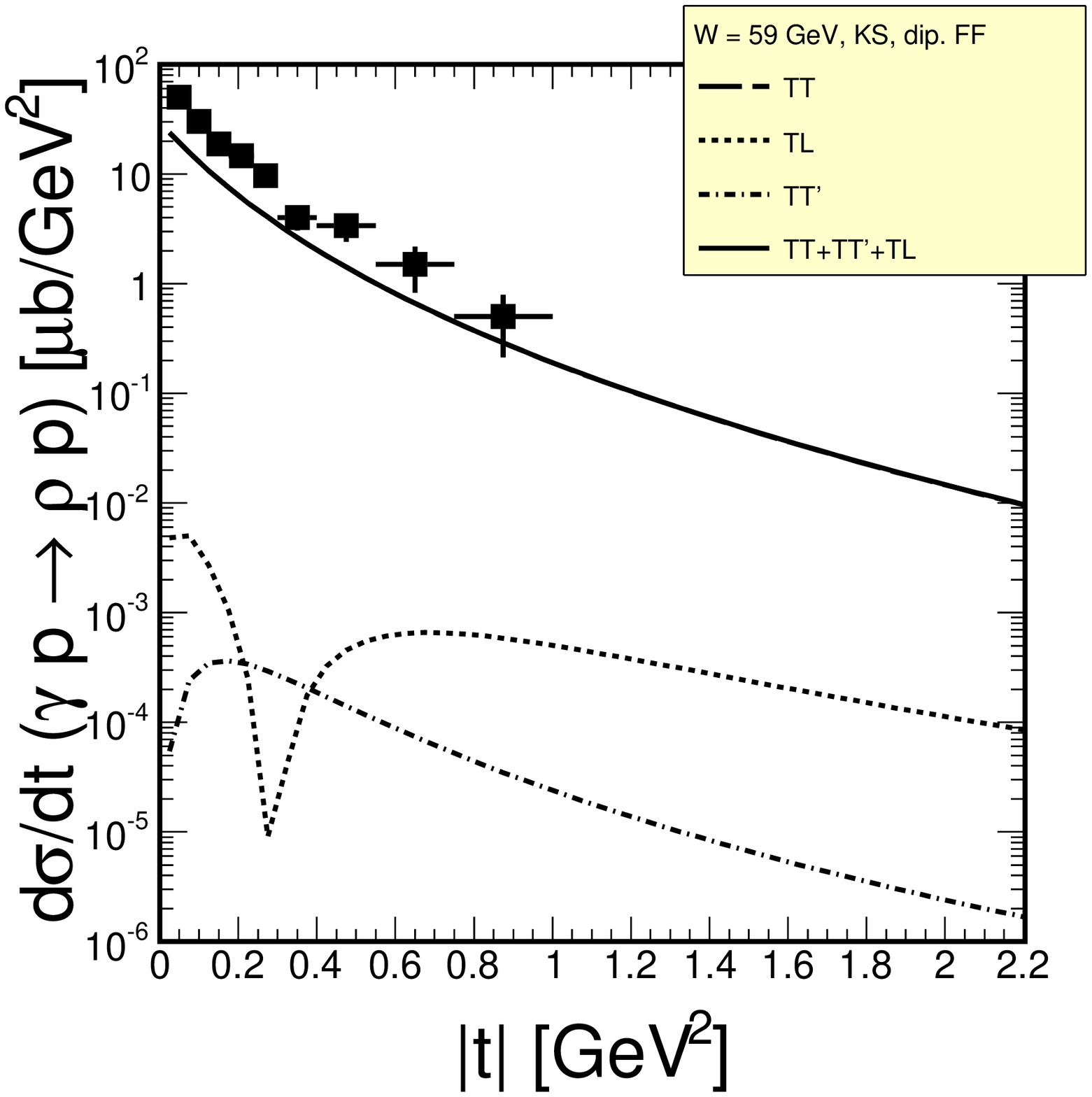}
\includegraphics[width=8.cm]{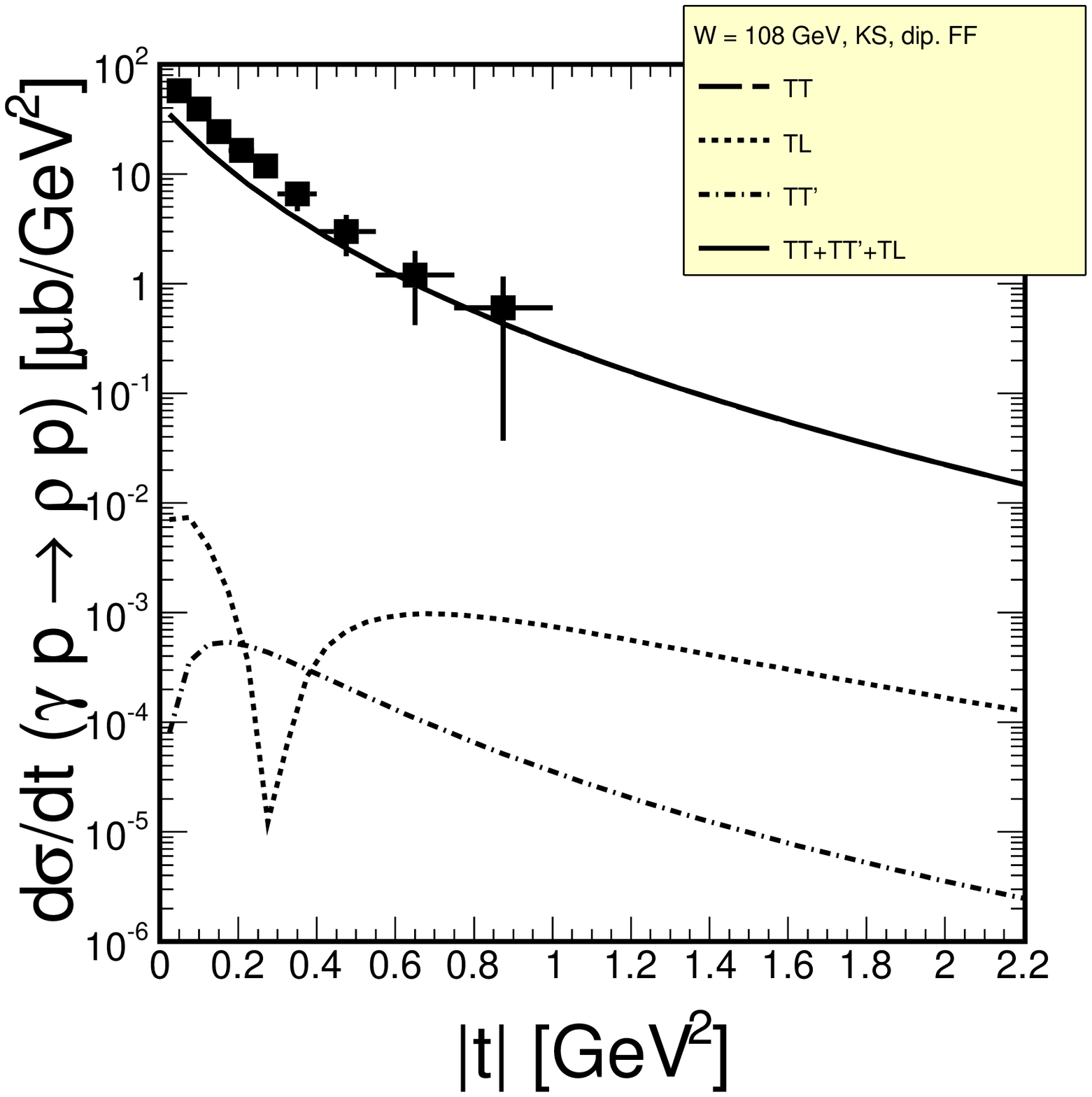}
\includegraphics[width=8.cm]{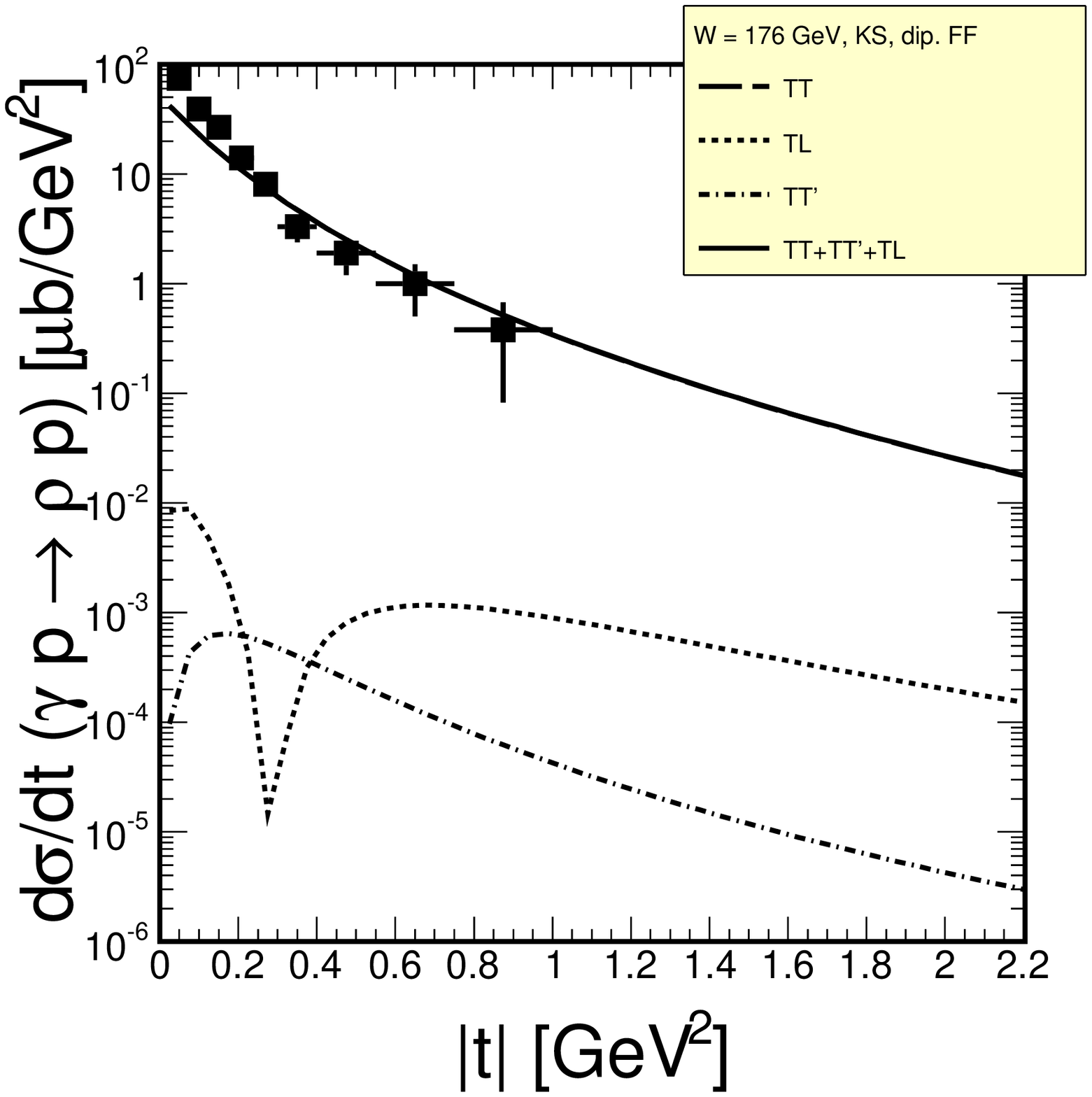}
\caption{Distribution in $t$ for different energies and
  the Kutak-Sta\'sto nonlinear UGD. Here the dipole parametrization of the form factor $G(\bDelta^2)$ was used. 
}
\label{fig:dsig_KS}
\end{figure}

In Fig.~\ref{fig:dsig_KS} the results for the Kutak-Sta\'sto UGD are
shown. Also here we observe no dip and a complete dominance of the
helicity conserving process. The CMS data are somewhat underpredicted with a slightly too hard $t$-dependence.

\begin{figure}[h]
\centering
\includegraphics[width=8.cm]{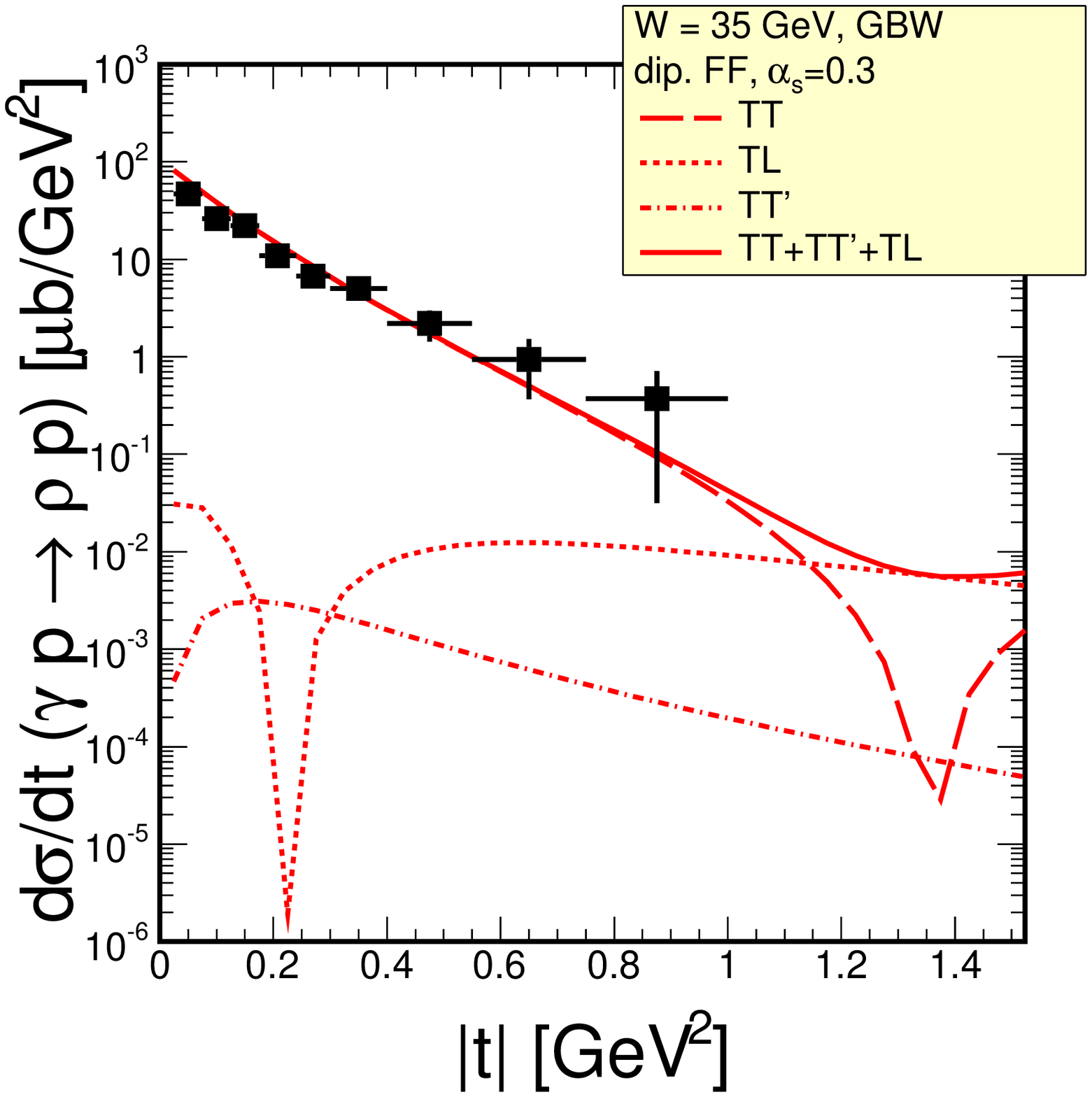}
\includegraphics[width=8.cm]{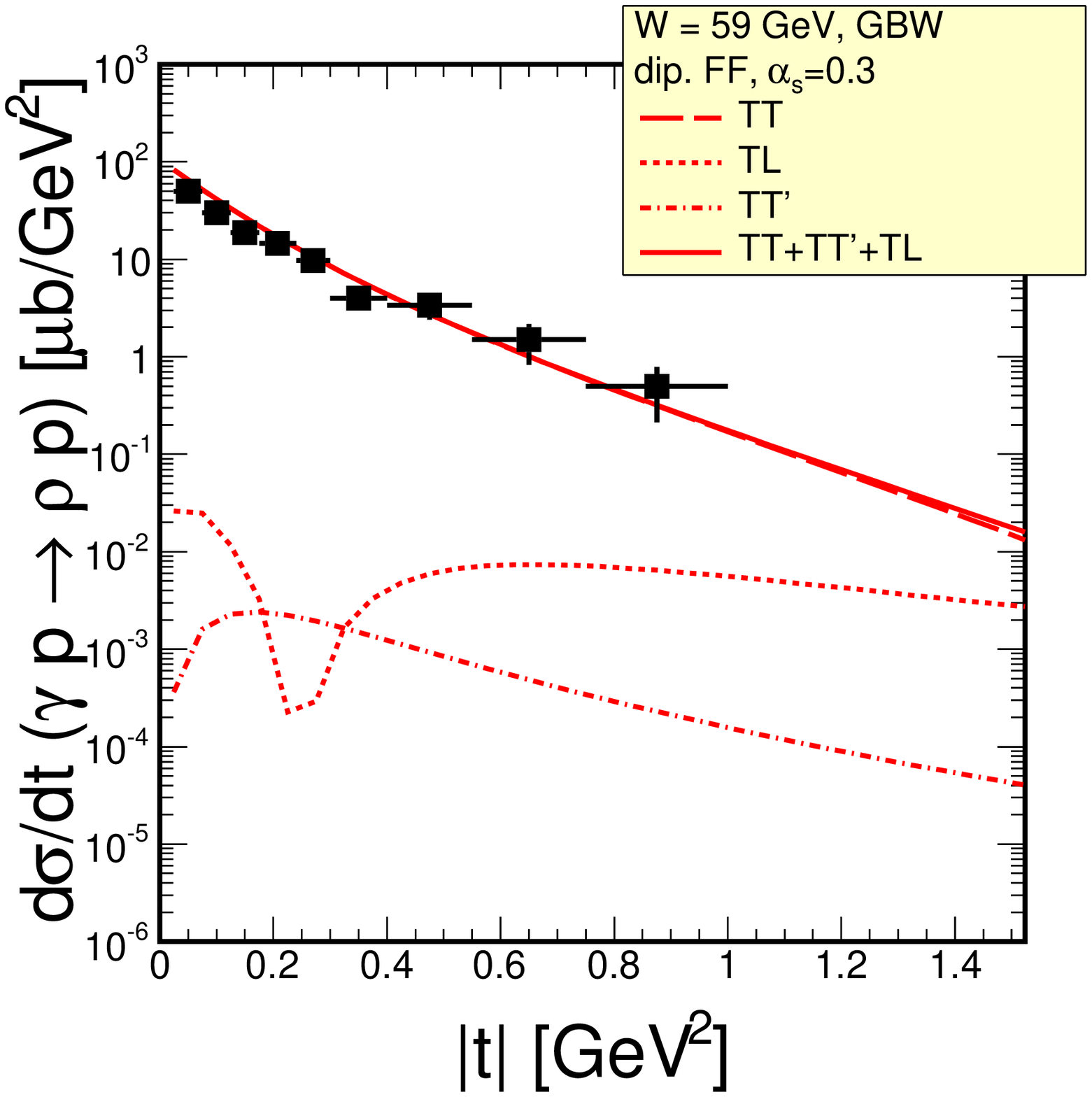}
\includegraphics[width=8.cm]{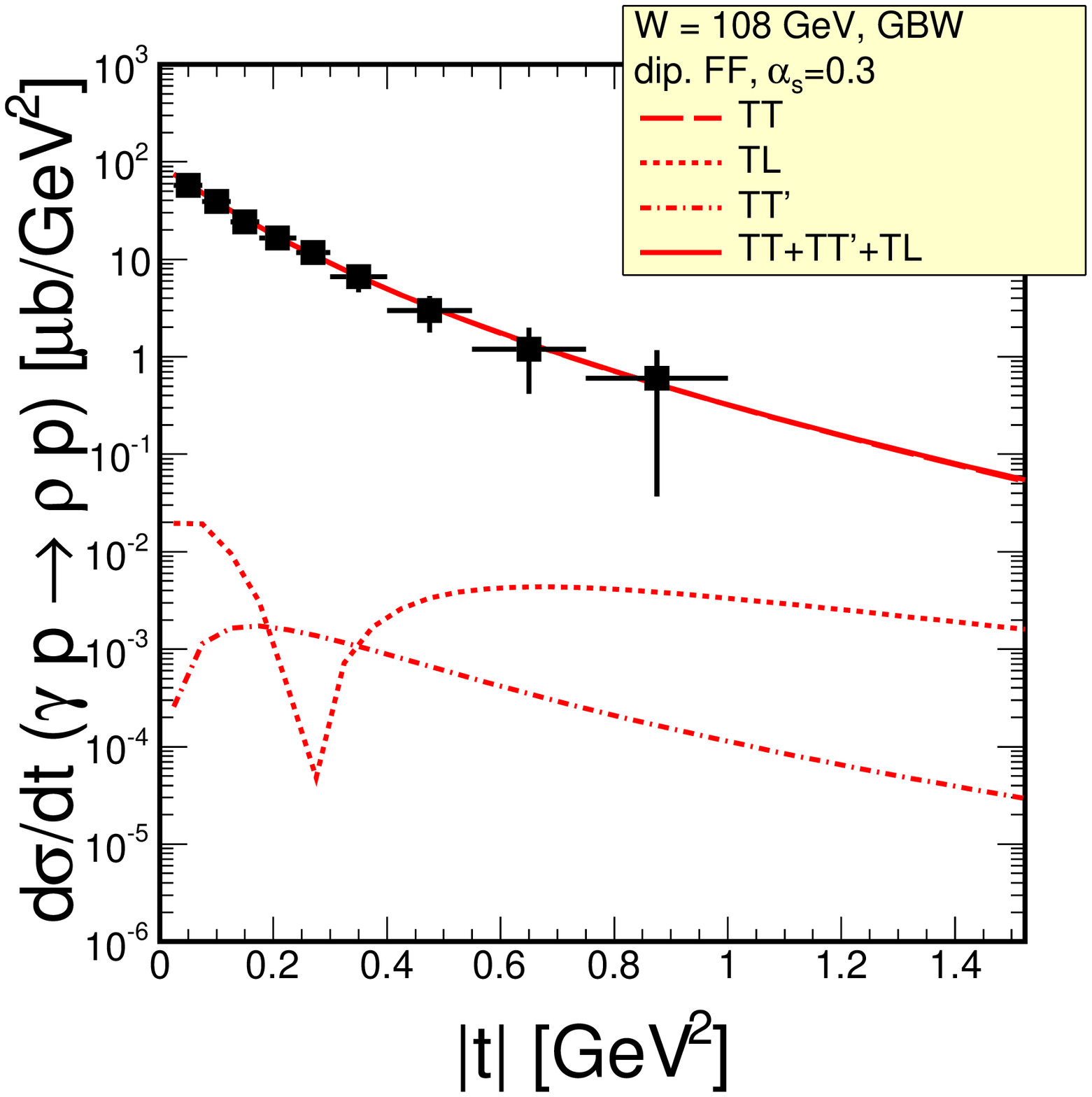}
\includegraphics[width=8.cm]{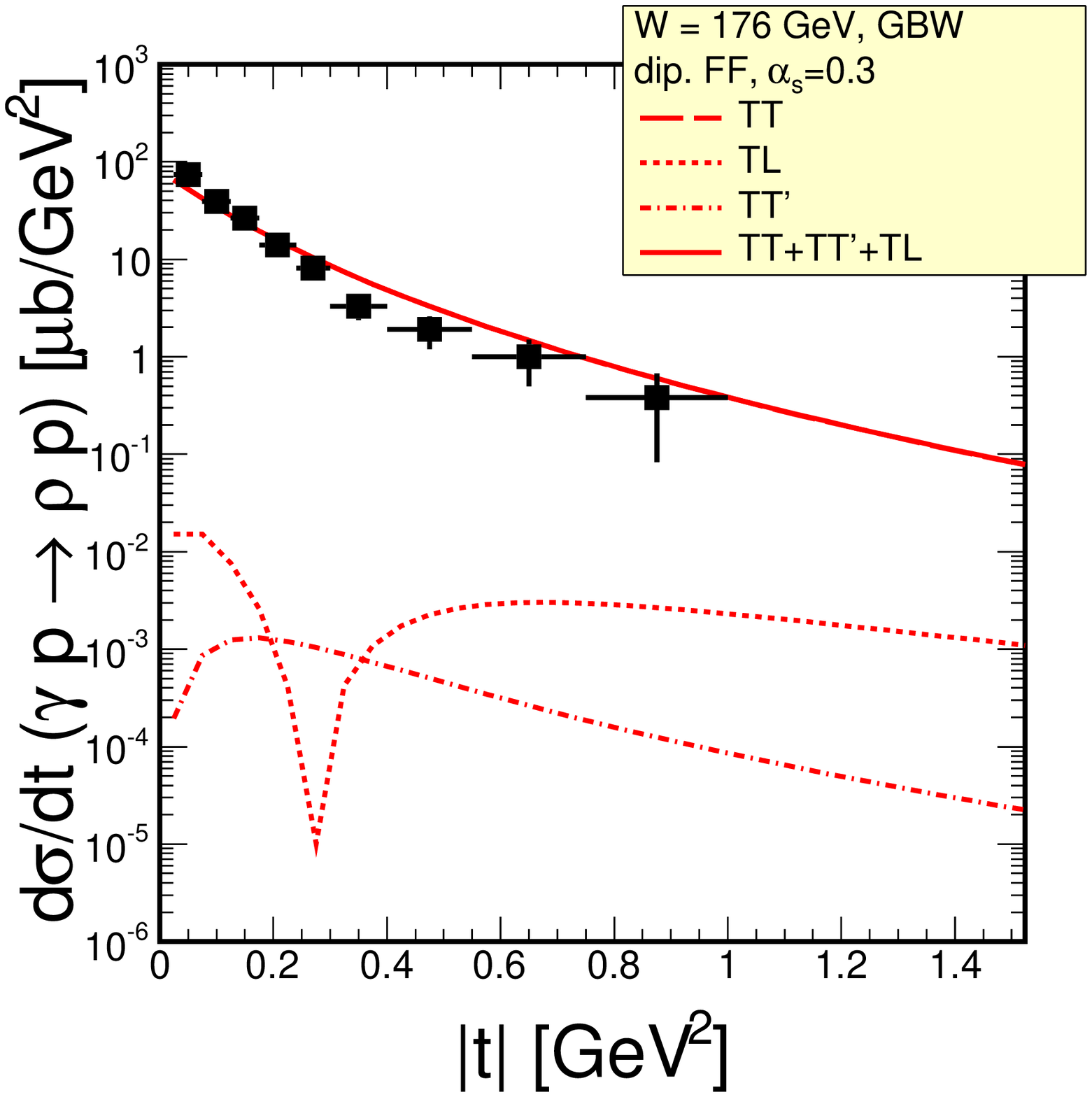}
\caption{Distribution in $t$ for different energies and the GBW UGD. 
Here the dipole parametrization of the form factor $G(\bDelta^2)$ was used.
}
\label{fig:dsig_GBW}
\end{figure}

The results for the GBW UGD are shown in Fig.~\ref{fig:dsig_GBW}. They
show generally the best agreement with the data. Also here, there is no dip within the measured region, and again helicity flip transitions are negligible.

\begin{figure}[h]
\centering
\includegraphics[width=8.cm]{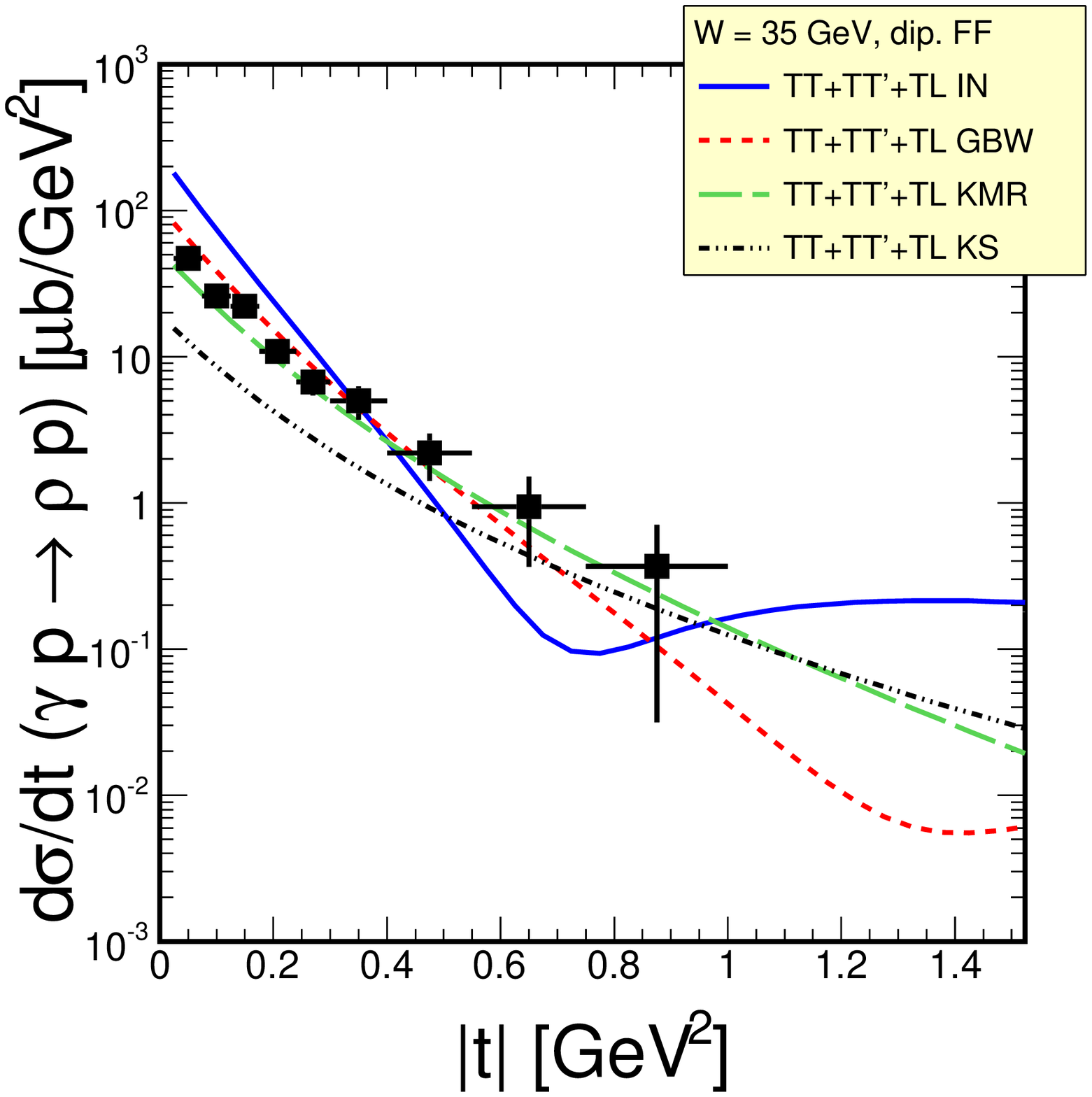}
\includegraphics[width=8.cm]{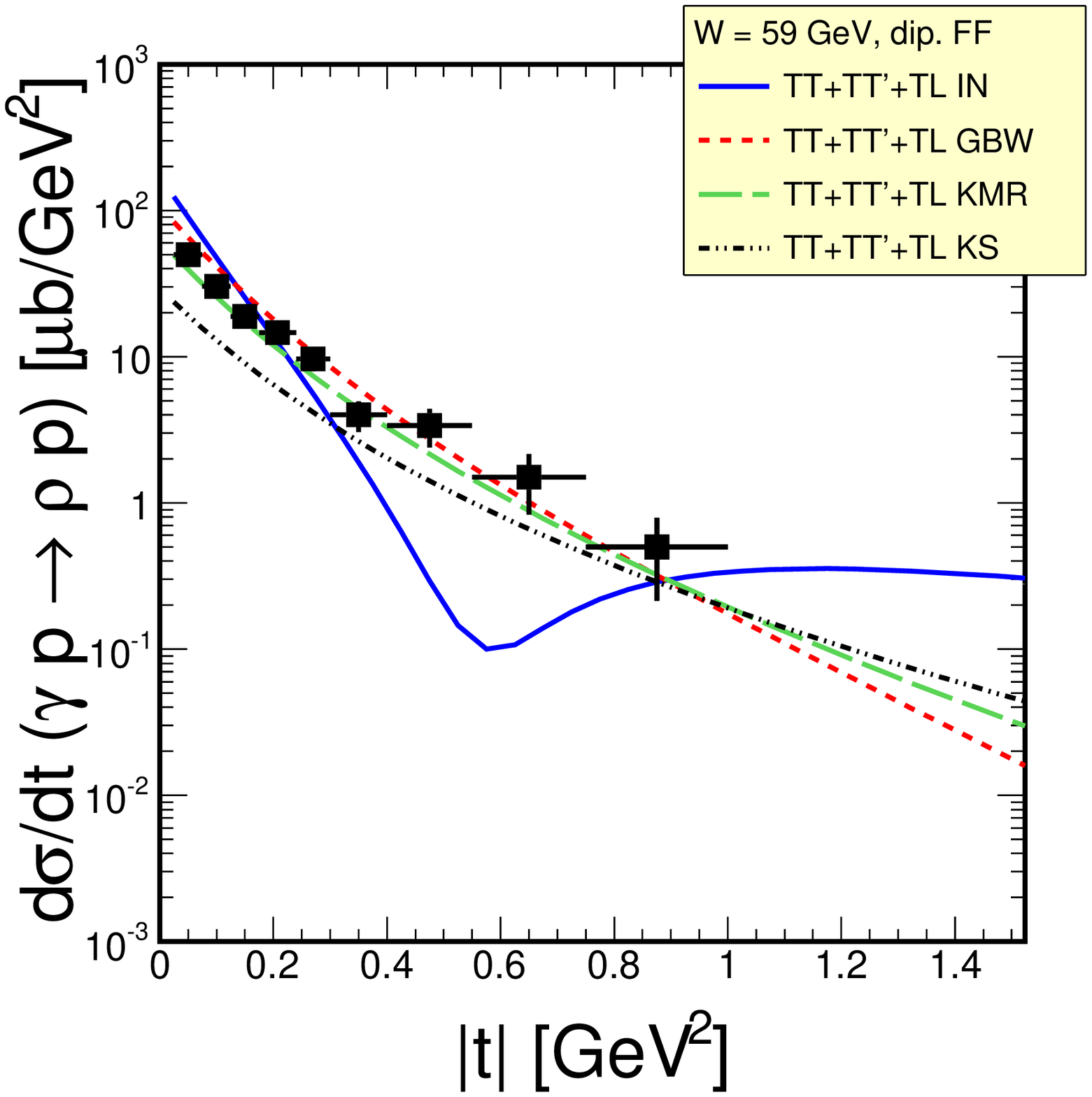}
\includegraphics[width=8.cm]{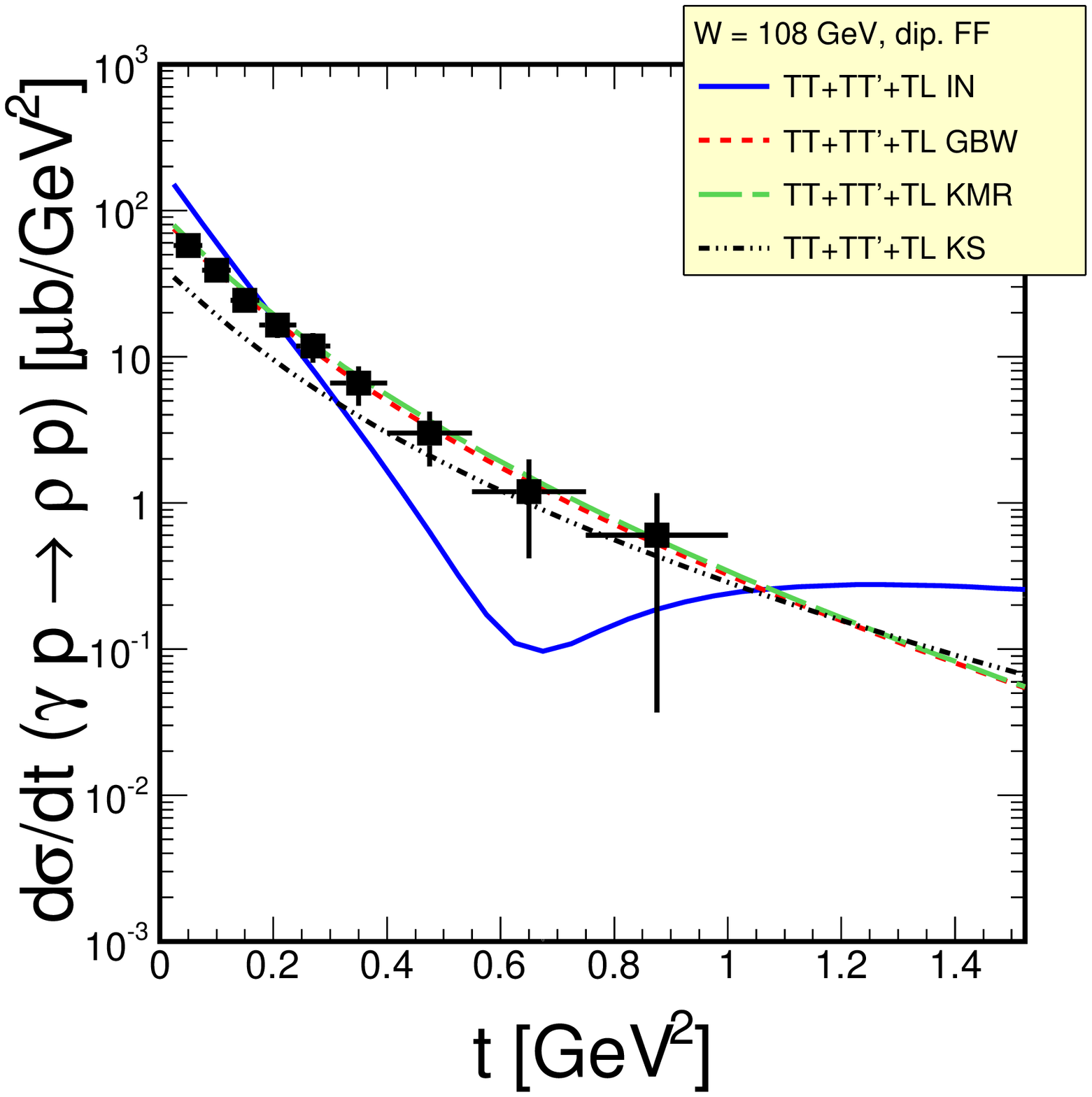}
\includegraphics[width=8.cm]{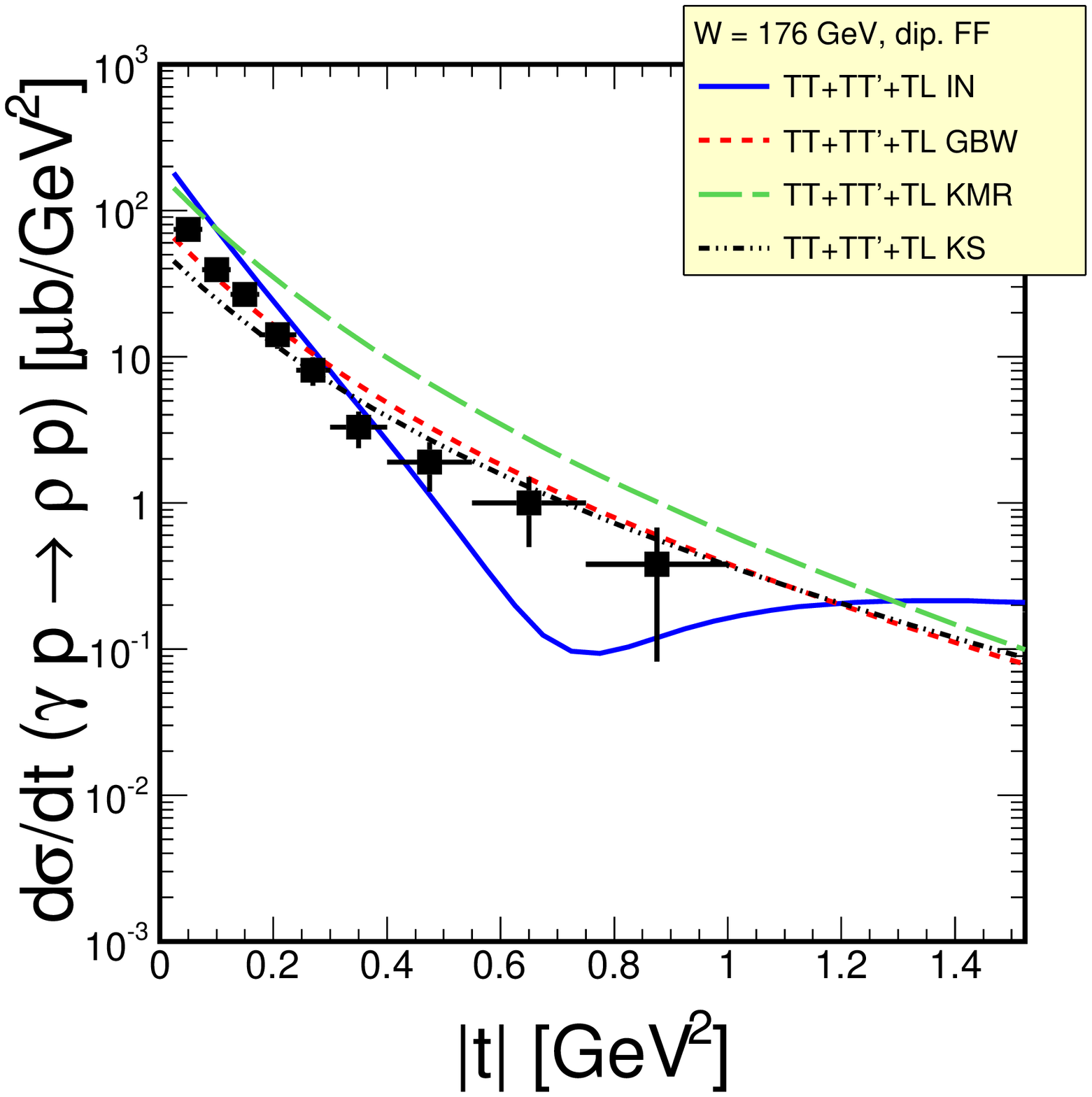}
\caption{Distribution in $t$ for different energies for the different UGDs.
}
\label{fig:dsig_all}
\end{figure}

Finally, we show for convenience the results for all UGDs summed over
all helicity combinations in Fig.~\ref{fig:dsig_all}. 

\section{Conclusions}

In this work we have investigated the differential cross section $d \sigma/dt$ for diffractive $\rho$-meson photoproduction at high energies in a model based on two-gluon exchange in the nonperturbative domain.
We have studied the role played by the often neglected helicity-flip amplitudes, which can contribute at finite $t$.

We have found that the large $|t|$-behaviour $d \sigma/dt$ depends on the form factor
describing the coupling of the pomeron to the $p \to p$ transition, while the dip-bump structure depends rather on the UGD used.

We have included traditional $T \to T$ contribution as well as somewhat
smaller $T \to L$ and $T \to T'$ (double spin-flip) contributions.
The relative amount  and differential shape of the subleading
contributions depends on the UGD used.

Some of the UGDs generate dips for $T \to T$ transition.
A good example is the Ivanov-Nikolaev UGD.
All UGDs generate dips for $T \to L$ transition.

Although helicity flip contributions are negligible for most of the models that we have employed,
they can be relevant, depending very much on the transverse momentum dependence of the UGD. This should be kept in mind especially when a specific model predicts diffractive dips and account only for the helcity conserving pieces, as is the common practice with most current models.

\section*{Acknowledgements}
This work was partially supported by the Polish National Science Center grant \\ UMO2018/31/B/ST2/03537 and by the Center for Innovation and Transfer of Natural Sciences
and Engineering Knowledge in Rzesz\'ow.

\bibliography{gammap_Vp}

\end{document}